\documentclass[reprint,amsmath,amssymb,prb,superscriptaddress,preprintnumbers,twocolumn]{revtex4-1}

\usepackage{graphicx}
\usepackage{hyperref}
\usepackage{xcolor}
\hypersetup{colorlinks,
	linkcolor={blue!50!black},
	citecolor={blue!50!black},
	urlcolor={blue!80!black}
}
\usepackage{color}

\begin{document}
\DeclareGraphicsExtensions{.pdf}

\title{Evidence for photoinduced sliding of the charge-order condensate in La$_{1.875}$Ba$_{0.125}$CuO$_4$}
\author {Matteo Mitrano\footnote{mmitrano@illinois.edu}}
\affiliation{Department of Physics and Frederick Seitz Materials Research Laboratory, University of Illinois, Urbana, IL 61801, USA}
\author {Sangjun Lee}
\affiliation{Department of Physics and Frederick Seitz Materials Research Laboratory, University of Illinois, Urbana, IL 61801, USA}
\author {Ali A. Husain}
\affiliation{Department of Physics and Frederick Seitz Materials Research Laboratory, University of Illinois, Urbana, IL 61801, USA}
\author {Minhui Zhu}
\affiliation{Department of Physics, University of Illinois, Urbana, IL 61801, USA}
\author {Gilberto de la Pe\~{n}a Munoz}
\affiliation{SLAC National Accelerator Laboratory, 2575 Sand Hill Road, Menlo Park, CA 94025, USA}
\author {Stella X. -L. Sun}
\affiliation{Department of Physics and Frederick Seitz Materials Research Laboratory, University of Illinois, Urbana, IL 61801, USA}
\author{Young Il Joe}
\affiliation{National Institute of Standards and Technology, Boulder, CO 80305, USA}
\author {Alexander H. Reid}
\affiliation{SLAC National Accelerator Laboratory, 2575 Sand Hill Road, Menlo Park, CA 94025, USA}
\author {Scott F. Wandel}
\affiliation{SLAC National Accelerator Laboratory, 2575 Sand Hill Road, Menlo Park, CA 94025, USA}
\author {Giacomo Coslovich}
\affiliation{SLAC National Accelerator Laboratory, 2575 Sand Hill Road, Menlo Park, CA 94025, USA}
\author {William Schlotter}
\affiliation{SLAC National Accelerator Laboratory, 2575 Sand Hill Road, Menlo Park, CA 94025, USA}
\author {Tim van Driel}
\affiliation{SLAC National Accelerator Laboratory, 2575 Sand Hill Road, Menlo Park, CA 94025, USA}
\author {John Schneeloch}
\affiliation{Condensed Matter Physics and Materials Science Department, Brookhaven National Laboratory, Upton, NY 11973, USA}
\author {G. D. Gu}
\affiliation{Condensed Matter Physics and Materials Science Department, Brookhaven National Laboratory, Upton, NY 11973, USA}
\author {Nigel Goldenfeld\footnote{nigel@illinois.edu}}
\affiliation{Department of Physics, University of Illinois, Urbana, IL 61801, USA}
\author {Peter Abbamonte\footnote{abbamonte@mrl.illinois.edu}}
\affiliation{Department of Physics and Frederick Seitz Materials Research Laboratory, University of Illinois, Urbana, IL 61801, USA}

\date{\today}

\begin{abstract}

We use femtosecond resonant soft x-ray scattering to measure the ultrafast optical melting of charge-order correlations in La$_{1.875}$Ba$_{0.125}$CuO$_4$. By analyzing both the energy-resolved and energy-integrated order parameter dynamics, we find evidence of a short-lived nonequilibrium state, whose features are compatible with a sliding charge density wave coherently set in motion by the pump. This transient state exhibits shifts in both the quasielastic line energy and its wave vector, as expected from a classical Doppler effect. The wave vector change is indeed found to directly follow the pump propagation direction. These results demonstrate the existence of sliding charge order behavior in an unconventional charge density wave system and underscore the power of ultrafast optical excitation as a tool to coherently manipulate electronic condensates.
\end{abstract}

\maketitle

\section{Introduction}
Charge carriers in quantum materials are often found to self-organize into ordered electronic phases, such as nematic states, Wigner crystals, superconductivity, charge-density waves (CDWs), spin density waves. Of particular interest are not only their emergent collective properties, but also their manipulation through a variety of control protocols (e.g. pressure, doping, temperature, and external fields) \cite{Keimer2015,Basov2017}.

When focusing on the response to external fields or currents, charge density waves (CDWs) represent a prototypical system for understanding the effects of these perturbations. Conventional CDWs are usually pinned to the lattice due to commensuration effects or to local impurities. If an applied electric field exceeds the pinning potential (i.e. for electric fields of a few V/cm), the CDW is depinned and free to slide with respect to the lattice \cite{Gruner,Gruner1988,Monceau2012}. This sliding behavior is accompanied by nonlinear transport signatures (e.g. non-Ohmic behavior of the current-voltage characteristics) and changes of the CDW periodicity, as documented in experiments on conventional CDW systems, such as NbSe$_3$ \cite{DiCarlo1993,Requardt1998,Danneau2002,Pinsolle2012}, K$_{0.3}$MoO$_3$ \cite{Tamegai1984,Hundley1989}, and elemental Cr \cite{Jacques2016}.

More recently, ultrafast lasers with peak electric fields in excess of several MV/cm have been applied to the study of charge ordered (CO) phases. They have enabled the observation of coherent collective order parameter dynamics following a sudden quench both with optical methods \cite{Torchinsky2013,Dakovski2015,Hinton2013} and x-ray scattering \cite{Huber2014,Beaud2014,Lee2012,Chuang2013}. In the specific case of copper oxides, near- and midinfrared pump pulses have been used to transiently enhance superconductivity above the equilibrium transition temperatures \cite{Fausti2011,Hu2014,Kaiser2014,Nicoletti2014} and simultaneously melt the CO correlations \cite{Forst2014,Khanna2016}. While ultrafast lasers can efficiently suppress the CO phase, it is not yet clear whether their peak fields are able to produce a coherent quasielastic dynamics of the condensate at ultrafast timescales.

To investigate this issue, we focus on the CO phase of La$_{1.875}$Ba$_{0.125}$CuO$_4$ (LBCO). With a correlation length reaching hundreds of lattice parameters, the incommensurate CO of this copper oxide superconductor is usually regarded as ``static'' \cite{Abbamonte2005,Hucker2011}, with a well-defined order parameter peak in momentum space. The crystal used here orders below T$_{CO}$ = 53 K, which coincides with an orthorhombic-to-tetragonal structural transition \cite{Tranquada2004,Abbamonte2005,Hucker2011}. 

Here, we use resonant time-resolved soft x-ray scattering, combined with both energy-resolved and energy-integrating detection, to analyze the in-plane pump-induced dynamics of CO in LBCO. We previously reported evidence of CO fluctuations at sub-meV energies \cite{Mitrano2018} by studying the relaxation behavior at long time delays. In this work we focus on transient quasielastic scattering shifts in both energy and momentum immediately after the pump arrival. We argue that these pump-induced changes are compatible with a Doppler effect caused by a field-induced sliding motion of the CO condensate.

\section{Experimental Details}

A high-quality boule of LBCO was grown by the floating zone method and cut into smaller single crystals \cite{Hucker2011}. The 2-mm-sized single crystal used in this study was cleaved in air in order to expose a fresh surface, miscut by 21 degrees with respect to the {\it ab} crystalline plane, and it was pre-oriented using a laboratory-based Cu K$\alpha$ x-ray source. The lattice parameters were determined to be a=b=3.787 \AA\ and c=13.23 \AA\ . The superconducting T$_c$, measured with a SQUID magnetometer, was found to be approximately 5 K.

\begin{figure}
    \centering
    \includegraphics[width=\columnwidth]{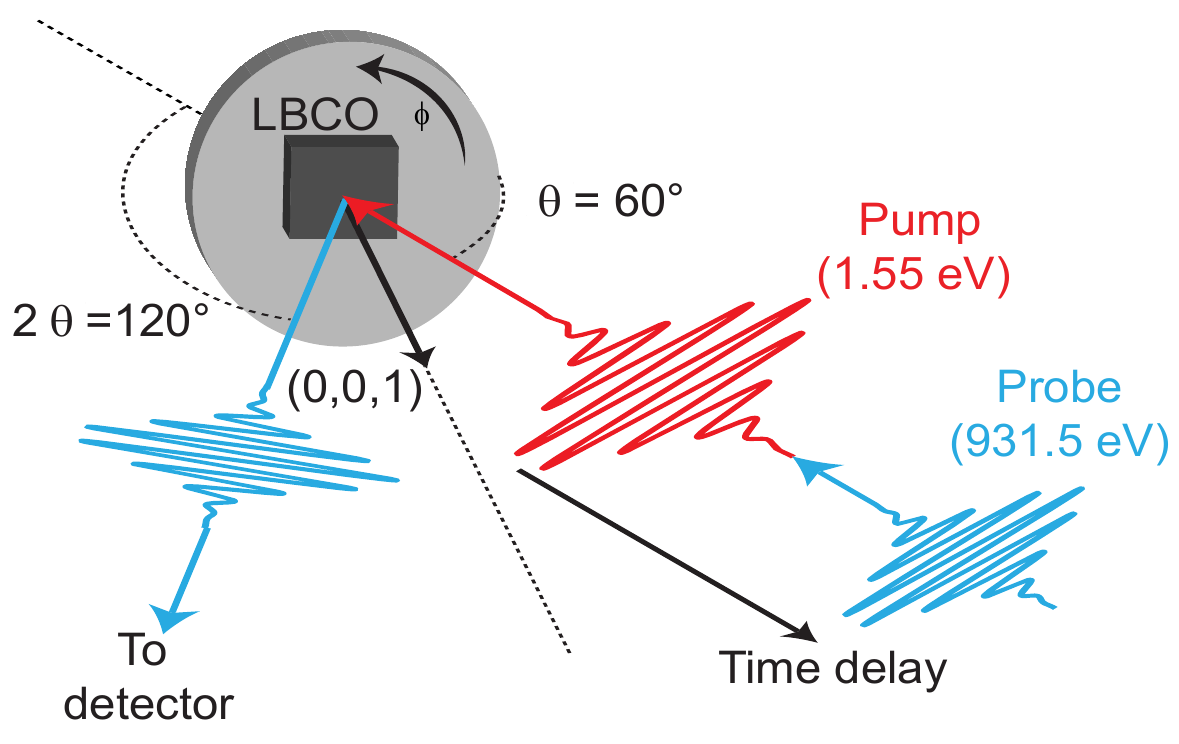} 
    \caption{(color online) Sketch of the experiment. The LBCO charge order is perturbed by 1.55 eV pump pulses and then probed by scattering of co-propagating soft x-ray FEL pulses resonantly tuned to the Cu L$_{3/2}$ edge. Both pump and probe pulses are p-polarized in the scattering plane. }
    \label{fig:exp_geometry_sketch}
\end{figure}

Optical pump, soft x-ray probe measurements have been performed at the Soft X-Ray (SXR) instrument of the Linac Coherent Light Source (LCLS) x-ray free electron laser (FEL) at the SLAC National Accelerator Laboratory (Menlo Park, USA) \cite{Schlotter2012}. Our measurements were carried out at a resonant soft x-ray scattering (RSXS) endstation \cite{Doering2011} in a $3\times10^{-9}$ Torr vacuum and at 12 K base temperature. The free electron laser was tuned to the Cu L$_{3/2}$ edge (931.5 eV) and with a 0.3 eV bandwidth after passing through a grating monochromator. The $p$ polarized x-ray pulses (60 fs duration,  1.5 $\mu$J energy, 120 Hz rep. rate) were focused down to a 1.5$\times$0.03 mm$^2$ elliptical spot on the sample. The 1.55 eV pump pulses (50 fs duration, 120 Hz rep. rate), also $p$ polarized, were generated with a Ti:sapphire amplifier and propagated colinearly with the x-rays into the RSXS endstation, as sketched in Fig. \ref{fig:exp_geometry_sketch}.
The pump was focused down to a 2.0$\times$1.0 mm$^2$ spot in order to probe a homogeneously excited sample volume. The shot-to-shot temporal pump-probe jitter was measured by means of a timing-tool \cite{Lemke2013,Harmand2013} and corrected by offline time sorting of the data. The overall time resolution was approximately 130 fs, as determined from the pump-probe crosscorrelation signal on a polished Ce:YAG crystal. Shot-to-shot intensity fluctuations from the FEL were corrected in the photodiode data through a reference intensity readout before the monochromator. 

The scattered x-rays were detected either by an energy-integrating avalanche photodiode (APD) on a rotating arm located 17.3 cm from the sample, or by a modular qRIXS grating spectrometer \cite{Chuang2017} mounted on a port at 135$^{\circ}$ with respect to the incident beam. The latter enabled energy-resolved measurements with a $\sim0.7$ eV energy resolution (FWHM) when using the second order of the grating. The spectrometer was equipped with an Andor CCD camera operated at 120 Hz readout rate in 1D binning mode along the nondispersive direction. Prior to further analysis and due to operation in a single-photon detection regime, we filtered the readout noise and retained only single- and double-photon events. The pump-probe time delay was controlled both electronically and through a mechanical translation stage.

\section{Results}
\subsection{Time-resolved RIXS data}

\begin{figure*}
    \centering
    \includegraphics[width=0.8\textwidth]{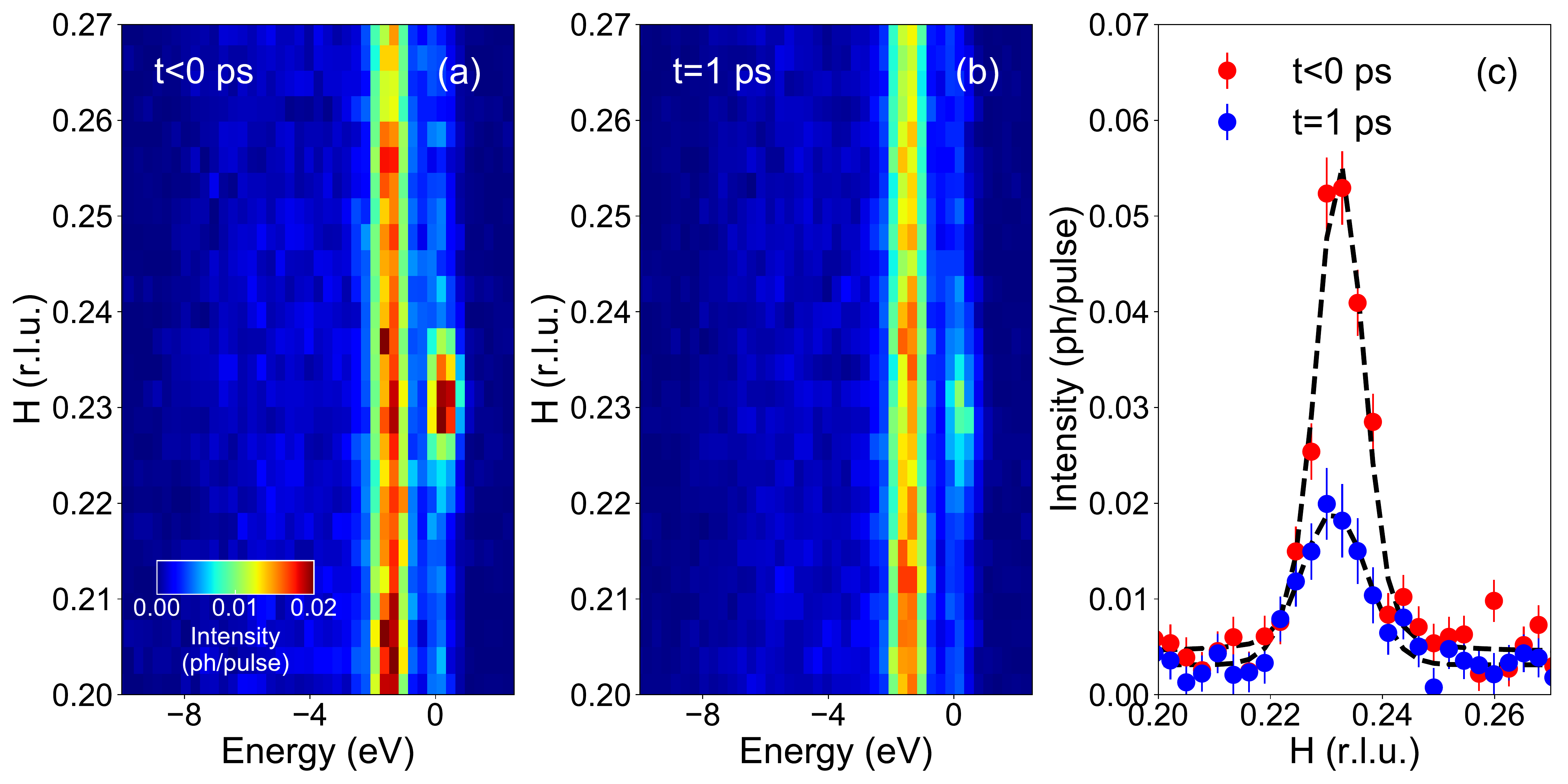} 
    \caption{(color online) tr-RIXS spectra as a function of momentum transfer along the $(H, 0, 1.5)$ r.l.u. direction (a) before, and (b) after the pump arrival. (c) Momentum dependence of the integrated elastic line (between -1.2 and 0.9 eV) before and after the pump arrival. Black dashed lines are pseudo-Voigt fits to the data. Error bars represent Poisson counting uncertainties.}
    \label{fig:trRIXS_theta_scan}
\end{figure*}

In order to investigate the pump-induced order parameter dynamics, we performed time-resolved Cu L$_{3/2}$ edge resonant inelastic x-ray scattering (tr-RIXS) measurements as a function of momentum and time delay.
First, we measured the inelastic LBCO spectra along the $(H, 0, 1.5)$ r.l.u. momentum direction for selected time delays [see Fig. \ref{fig:trRIXS_theta_scan}]. In this work, we use the $(H, K, L)$ Miller indices in the tetragonal notation to define the peak positions in momentum space. Panel (a) shows three distinct features of the sample spectrum at equilibrium: two momentum-independent components at -1.7 and -6.0 eV, which are the $dd/Cu^{2+}$ and charge-transfer (CT) excitations respectively \cite{Ghiringhelli2004}, and a sharp elastic peak located at $H=0.23$, corresponding to the order parameter of the CO phase.

As the system is excited with approximately 0.1 mJ/cm$^2$ pump pulses [corresponding to a peak electric field of approximately 0.61 MV/cm, see Fig. \ref{fig:trRIXS_theta_scan}(b)], its intensity decreases by 60\%, in line with previous pump-induced charge-order melting measurements \cite{Khanna2016,Forst2014}. The inelastic features appear mostly unperturbed.
By integrating the quasielastic peak in a 2.1 eV energy window around the zero energy loss [see Fig. \ref{fig:trRIXS_theta_scan}(c)], we are able to separate the order parameter from the inelastic background and to quantitatively assess pump-induced lineshape changes. The momentum-dependent elastic peak is well described at each time delay $t$ by a pseudo-Voigt profile
\begin{equation}
\begin{aligned}
I(q)\bigg|_{t}={}&(I_0+mq)+f\frac{1}{\pi g}\frac{A}{1+\bigg(\frac{q-Q_{CO}}{g}\bigg)^2}\\
&+(1-f)\frac{A}{g}\sqrt{\frac{\ln2}{\pi}}\exp\bigg[-\ln2\bigg(\frac{q-Q_{CO}}{g}\bigg)^2\bigg],
\end{aligned}
\label{eq:pseudovoigt}
\end{equation}
where the first term represents a linear background, while the second and third terms represent a Lorentzian and Gaussian with linear mixing parameter $f$. The last two terms share the same amplitude $A$ and FWHM $2g$. 

A fit to the energy-integrated elastic peak reveals that the FWHM broadens from $(9.96\pm0.58)\times10^{-3}$ r.l.u. to $(13.8\pm1.2)\times10^{-3}$ r.l.u. The 40$\%$ broadening of the elastic line implies that the pump generates defects (e.g. dislocations) in the ordered phase, thus reducing the correlation length $\xi=1/g$ from 201 \AA\  to 145 \AA\ . This defect-mediated melting is qualitatively different from what was observed in charge-ordered ${\mathrm{La}}_{1.75}{\mathrm{Sr}}_{0.25}{\mathrm{NiO}}_{4}$ \cite{Lee2012,Chuang2013}, thus suggesting different excitation pathways for 214 nickelates and cuprates. The time-dependent evolution of these defects obeys dynamical critical scaling, as discussed in Ref. \cite{Mitrano2018}.

\begin{figure*}
    \centering
    \includegraphics[width=0.8\textwidth]{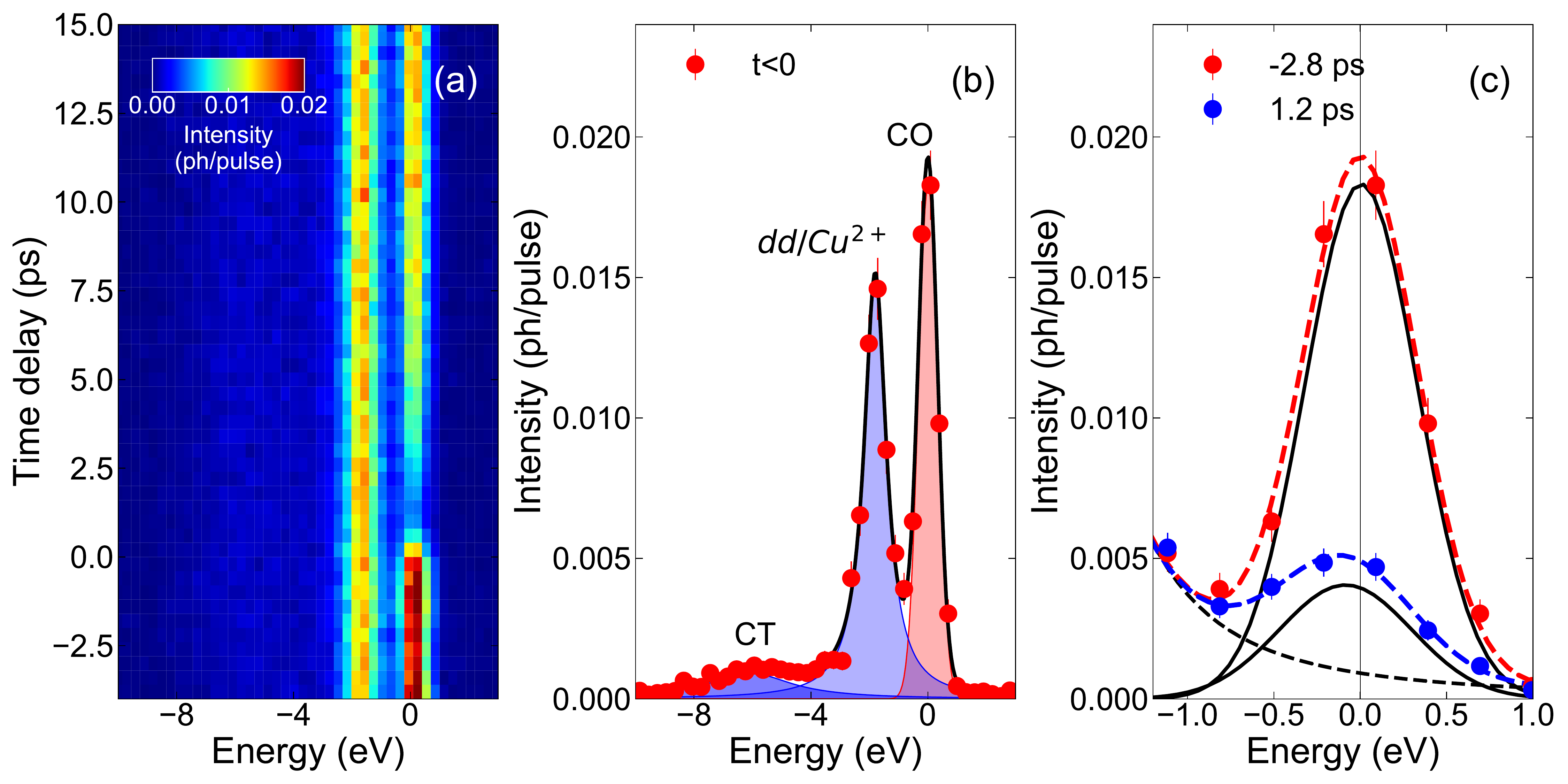}
    \caption{(color online) (a) tr-RIXS spectra at $Q_{CO}=(0.23,0.00,1.50)$ r.l.u. as a function of pump-probe time delay. The data are binned in 400 fs time steps to improve statistics. (b) Fit components of the tr-RIXS spectra at $Q_{CO}=(0.23,0.00,1.50)$ r.l.u. The shaded red area is the Gaussian used to model the quasielastic scattering, while the two shaded blue areas are the two Lorentzians capturing the $dd$ and CT excitations respectively. (c) Zoom of the quasielastic scattering before (red) and after (blue) the pump arrival. Data are shown as filled circles, while the red and blue dashed lines are fits to the data. The solid black lines are the Gaussian components of the fit, and the black dashed line is the Lorentzian fit to the dd excitations. Error bars are Poisson counting uncertainties.}
    \label{fig:trRIXS_map_fit}
\end{figure*}

After being suppressed, the order parameter relaxes back to equilibrium. tr-RIXS spectra at $Q_{CO}=(0.23,0.00,1.50)$ r.l.u. were acquired for a dense distribution of time delays [Fig. \ref{fig:trRIXS_map_fit}a], then time sorted, and rebinned offline in 400-fs steps to improve statistics. The entire dataset has been acquired in approximately 200 minutes at full (monochromatized) FEL beam.

The time-dependent RIXS spectra confirm that the pump only affects the quasielastic scattering associated with the charge density modulation, without perturbing $dd$ or CT excitations. We fit the data to a three-component function, independently for each time delay. The CT excitations and the $dd$ excitations are described by two Lorentzians, while the elastic line is well captured by a Gaussian, since its energy width is dominated by the intrinsic 0.7 eV spectrometer resolution [Fig. \ref{fig:trRIXS_map_fit}(b)]. We note here that an additional 30-meV energy broadening of the spectral features due to the short x-ray pulse duration is negligible when compared to the instrument resolution.

\begin{figure}
    \centering
    \includegraphics[width=\columnwidth]{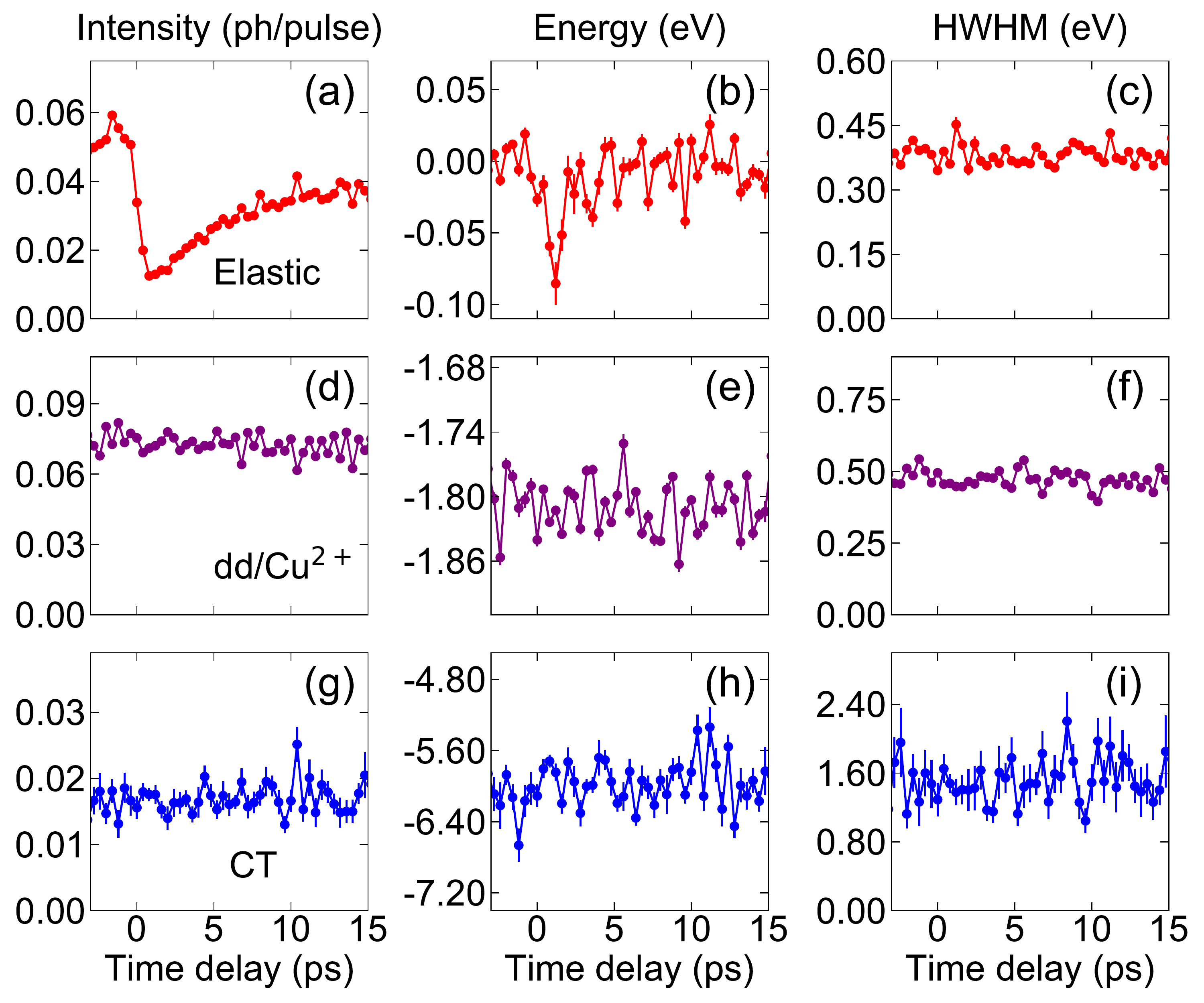}
    \caption{(color online) Fit parameters for the RIXS spectra at $Q_{CO}=(0.23,0.00,1.50)$ r.l.u. as a function of pump-probe time delay. Panels (a)-(c) respectively show the integrated intensity, energy, and HWHM of the CO elastic line. Panels (d)-(f) report the same parameters for the peak associated to the dd excitations/Cu$^{2+}$ fluorescence line. Panels (g)-(i) describe instead the spectral feature associated to the charge-transfer excitations. Error bars represent fit uncertainties.}
    \label{fig:trRIXS_fit_parameters}
\end{figure}

The time-dependent fits reveal that the order parameter is not only suppressed by the pump [Fig. \ref{fig:trRIXS_fit_parameters}(a)], but also exhibits a short-lived, 80 meV pump-induced redshift along the energy loss axis [Fig. \ref{fig:trRIXS_fit_parameters}(b)]. The intensity, peak energy and width of $dd$ [Fig. \ref{fig:trRIXS_fit_parameters}(d)-\ref{fig:trRIXS_fit_parameters}(f)] and CT excitations [Fig. \ref{fig:trRIXS_fit_parameters}(g)-\ref{fig:trRIXS_fit_parameters}(i)] do not exhibit visible time-dependent behavior within the fit uncertainties. The tail of the $dd$/Cu$^{2+}$ peak did not influence the elastic peak position deduced from the fits (see Appendix for further discussion). Due to the limited energy resolution, we cannot conclude whether this phenomenon [visible in the raw data shown in Fig.  \ref{fig:trRIXS_map_fit}(c)] is due to an actual redshift of the elastic line or to the creation of low-energy ($\hbar\omega<0.3$ eV) inelastic excitations.

\subsection{Evidence of photoinduced sliding}
\begin{figure*}
    \centering
    \includegraphics[width=0.8\textwidth]{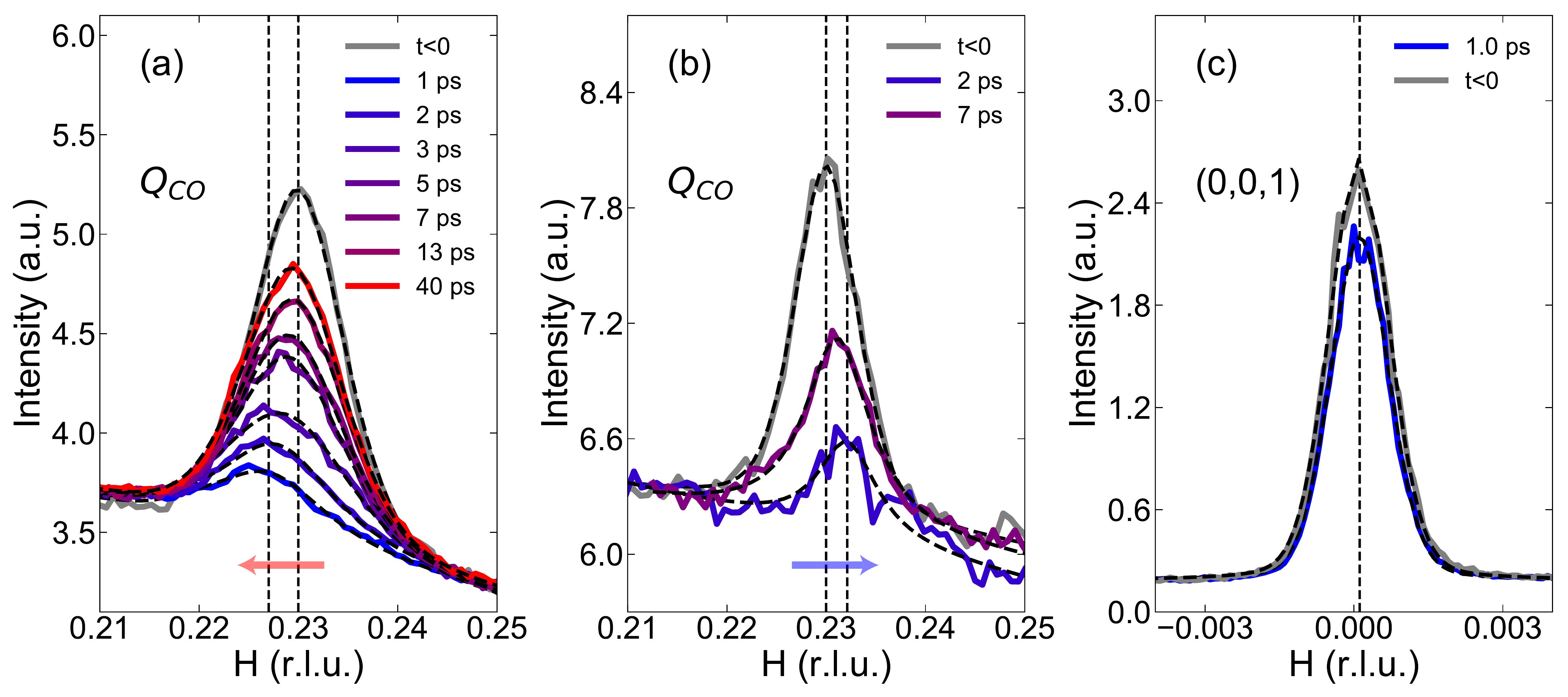}
    \caption{(color online) Transverse momentum scan of the charge order peak along the $(H,0.00,1.50)$ r.l.u. direction for a selection of time delays. Dashed lines are fits using eq. \ref{eq:pseudovoigt}. Dashed vertical lines mark the position of the peak at equilibrium and for the maximum observed shift. Panel (a) shows data acquired at $\phi=0$, while panel (b) at $\phi\sim\pi$. Note that the change in momentum reverses sign between the two configurations. (c) Transverse momentum scan along the (H,0,1) direction of the structural (0,0,1) Bragg peak before and after the pump arrival. The vertical dashed line marks the peak position before and after the pump.}
    \label{fig:Fig_recoil_PRB_bkg}
\end{figure*}

In addition to the energy redshift, the CO peak also changes in momentum, as revealed by the tr-RIXS data in Fig. \ref{fig:trRIXS_theta_scan}(c). More quantitatively, the integrated quasielastic peak moves by $\Delta q=Q_{CO}(t=1\:\textrm{ps})-Q_{CO}(t<0)=-(1.11\pm0.49)\times10^{-3}$ r.l.u. along the $(H,0,1.5)$ direction. To gain further insight, we repeat the measurement with an energy-integrating APD, which enables an increased signal-to-noise ratio [see Fig. \ref{fig:Fig_recoil_PRB_bkg}(a)].
Here we find a maximum pump-induced momentum shift $\Delta q\sim0.003$ r.l.u., which is significantly larger than the current-induced shifts reported in NbSe$_3$ ($\Delta q\sim0.0006$ r.l.u.)  \cite{DiCarlo1993,Pinsolle2012}, and similar in magnitude to laser-induced diffraction results in elemental Cr ($\Delta q\sim0.002$ r.l.u.)\cite{Jacques2016}.
This shift occurs in the scattering plane, along the H momentum direction, but not along the perpendicular K direction \cite{Mitrano2018}. 

\begin{figure}
    \centering
    \includegraphics[width=\columnwidth]{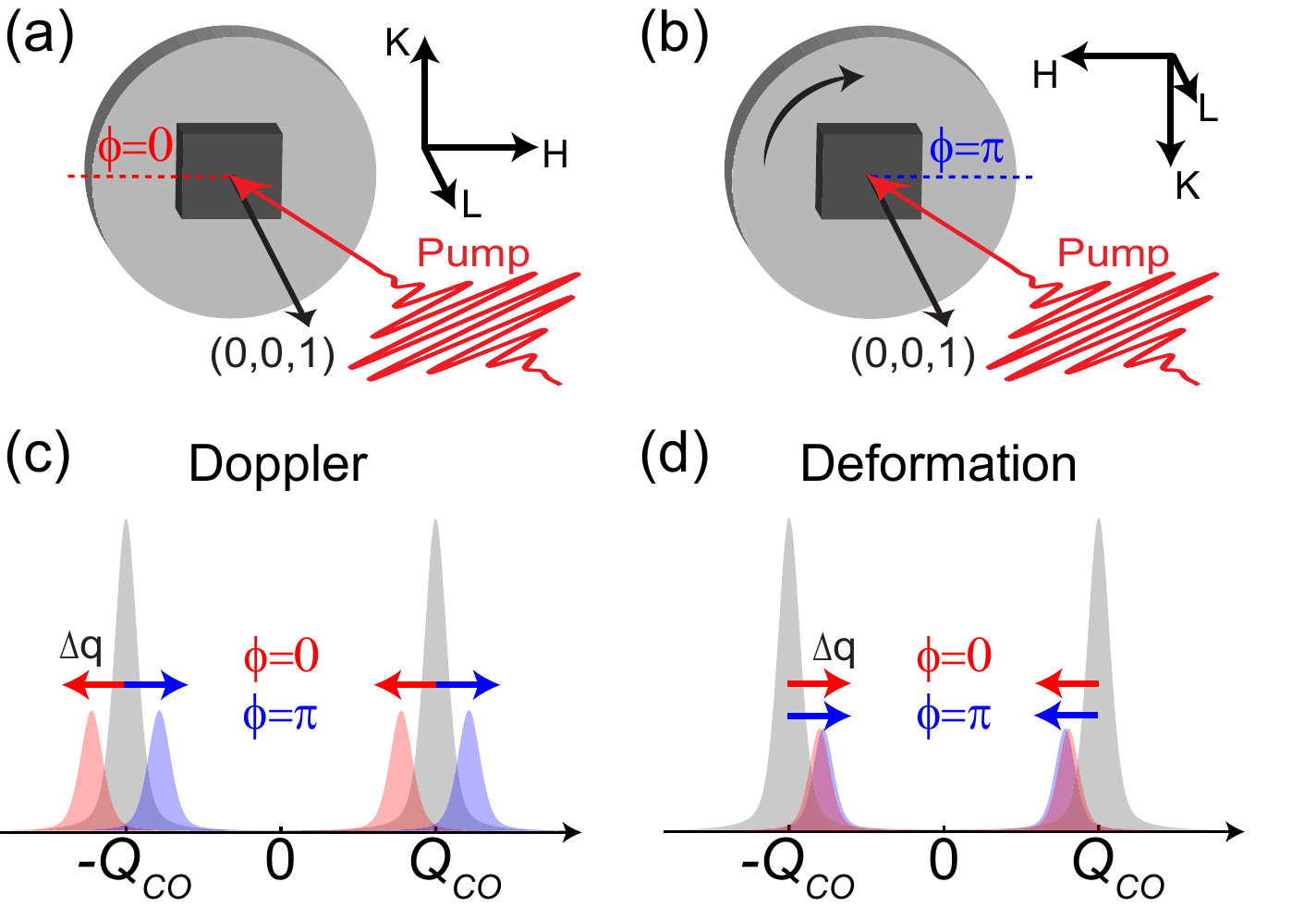}
    \caption{(color online) Azimuthal rotation of the LBCO crystal. The $Q_{CO} =(0.23,0.00,1.50)$ r.l.u.  diffraction peak has been measured in two configurations: (a) $\phi=0$, with the pump propagating from positive to negative H, and (b) $\phi=\pi$, from negative to positive H. (c) Observed Doppler recoil of the condensate from the equilibrium diffraction peak (gray) to the $\Delta q$-shifted reflection (red, blue). (d) Diffraction peak shift due to a change of the charge density periodicity (deformation). Red and blue arrows indicate the direction of the shift for $\phi=0,\:\pi$, respectively.}
    \label{fig:shift_interpretation}
\end{figure}

This pump-induced phenomenon could originate from any of three possible effects: (1) a change in the sample refractive index in the soft x-ray regime, which would alter the measured Bragg reflection angle \cite{Smadici2013,Achkar2016}, (2) a change in the CO periodicity, or (3) a collective recoil of the condensate.

A pump-induced change in the refractive index would also affect other diffraction peaks. However, we observe no corresponding pump-induced shift in the $(0,0,1)$ Bragg reflection of the low-temperature tetragonal structure [see Fig. \ref{fig:Fig_recoil_PRB_bkg}(c)], thus ruling out this first possible explanation.

On the other hand, changes in the charge order periodicity have been observed in a wide variety of systems under the effect of external electric fields or currents \cite{DiCarlo1993,Pinsolle2012,Jacques2016} and are ascribed to deformations of the density wave modulation $\rho(x)$, defined as
\begin{equation}
\rho(x)=\rho_0+\Delta\rho\cos(Q_{CO}x+\varphi),
\end{equation}
where $\rho_0$ is the background charge density, $\Delta\rho$ is the CO modulation, $x$ is the spatial coordinate, and $\varphi$ is the CO phase. Deforming the charge order modulation is equivalent to introducing a spatial dependence in the phase $\varphi$. To the lowest order, a linear gradient in the phase $\varphi(x)=\varphi_0+(\partial_x\varphi)x$ is equivalent to altering the charge order wave vector. In other words, the distorted charge density modulation is likely to have the form 
\begin{equation}
\begin{aligned}
\rho'(x)={}&\rho_0+\Delta\rho\cos(Q_{CO}x+\varphi_0+(\partial_x\varphi)x)\\
&=\rho_0+\Delta\rho\cos(Q'_{CO}x+\varphi_0)
\end{aligned}
\end{equation}
with $Q'_{CO}=Q_{CO}+(\partial_x\varphi)$. 

In order to test this possibility, we rotated the sample azimuthal angle $\phi$ by $180^{\circ}$ and repeated the measurement at the same wave vector $Q_{CO} =(0.23,0.00,1.50)$ r.l.u. under the same excitation conditions [see Fig. \ref{fig:shift_interpretation}(a)-(b)]. This is equivalent to changing the orientation of the pump propagation axis with respect to the H crystallographic direction. In our geometry $\phi=0$ corresponds to a pump propagating from positive to negative H values, while $\phi=\pi$ from negative to positive H.
If the shift were due to a periodicity change, such a rotation would not affect the $\Delta q$ momentum shift as measured in the reference frame of the sample and it would be independent of the pump propagation direction. Furthermore, the CO reflection located at $-Q_{CO}$ would shift in the opposite direction, as shown in Fig. \ref{fig:shift_interpretation}d.\\
Surprisingly, we instead observe a reversal in the momentum shift [Figs. \ref{fig:shift_interpretation}(c) and \ref{fig:Fig_recoil_PRB_bkg}(b)], meaning the peak moves in a fixed direction with respect to the propagation of the pump, {\it not the crystal axes}, excluding a CO periodicity change, either uniform or non-uniform. In other words, the pump exchanges momentum $\Delta q$ with the condensate along its propagation axis. In this scenario, the two reflection at $\pm Q_{CO}$ are expected to move in the same pump-dependent momentum direction [see Fig. \ref{fig:shift_interpretation}(c)].

This reversal in momentum shift leads to the conclusion that the pump induces a sliding motion of the CO condensate with a uniform momentum $\Delta q$ or, equivalently, creates a nonequilibrium population of collective modes exhibiting a nonzero center-of-mass momentum. If the condensate is coherently set in motion, the elastic peak located at $(Q_{CO},\omega=0)$ will shift both in energy and momentum $(Q_{CO}\pm\Delta q,\omega\mp v\Delta q)$ \cite{Coppersmith1984} and the direction will be dictated by the applied field. This is, in essence, a manifestation of the Doppler effect for a moving electronic system. 

As this phenomenon is expected to exhibit a simultaneous shift of the elastic line by $\Delta\omega=\mp v\Delta q$ \cite{Coppersmith1984}, where $v$ is the phason velocity of the CDW, one could naively identify the observed 80-meV RIXS redshift in Figs. \ref{fig:trRIXS_map_fit}(c) and \ref{fig:trRIXS_fit_parameters}(b) with such energy change. This would imply a condensate velocity $v=\Delta\omega /\Delta q=16$ eV$\times$\AA\ ($v=2.4\times10^{6}$ m/s), far greater than the nodal Fermi velocity $v_F\sim1.7$ eV$\times$\AA\ ($2.6\times10^{5}$ m/s) \cite{Valla2006}, which is regarded as a limiting value for the electronic velocity of a CDW \cite{Gruner}. Given this rather unphysical value, we conclude that the observed energy redshift, characterizing the spectrum of the sliding CO, is due to a distinct effect coexisting with the Doppler shift. For example, it could be that the high-energy excitation spectrum of a sliding CDW is different from that of a CDW in equilibrium. Higher resolution RIXS measurements with next-generation FEL instrumentation are needed to determine its origin.
\begin{figure}
    \centering
    \includegraphics[width=\columnwidth]{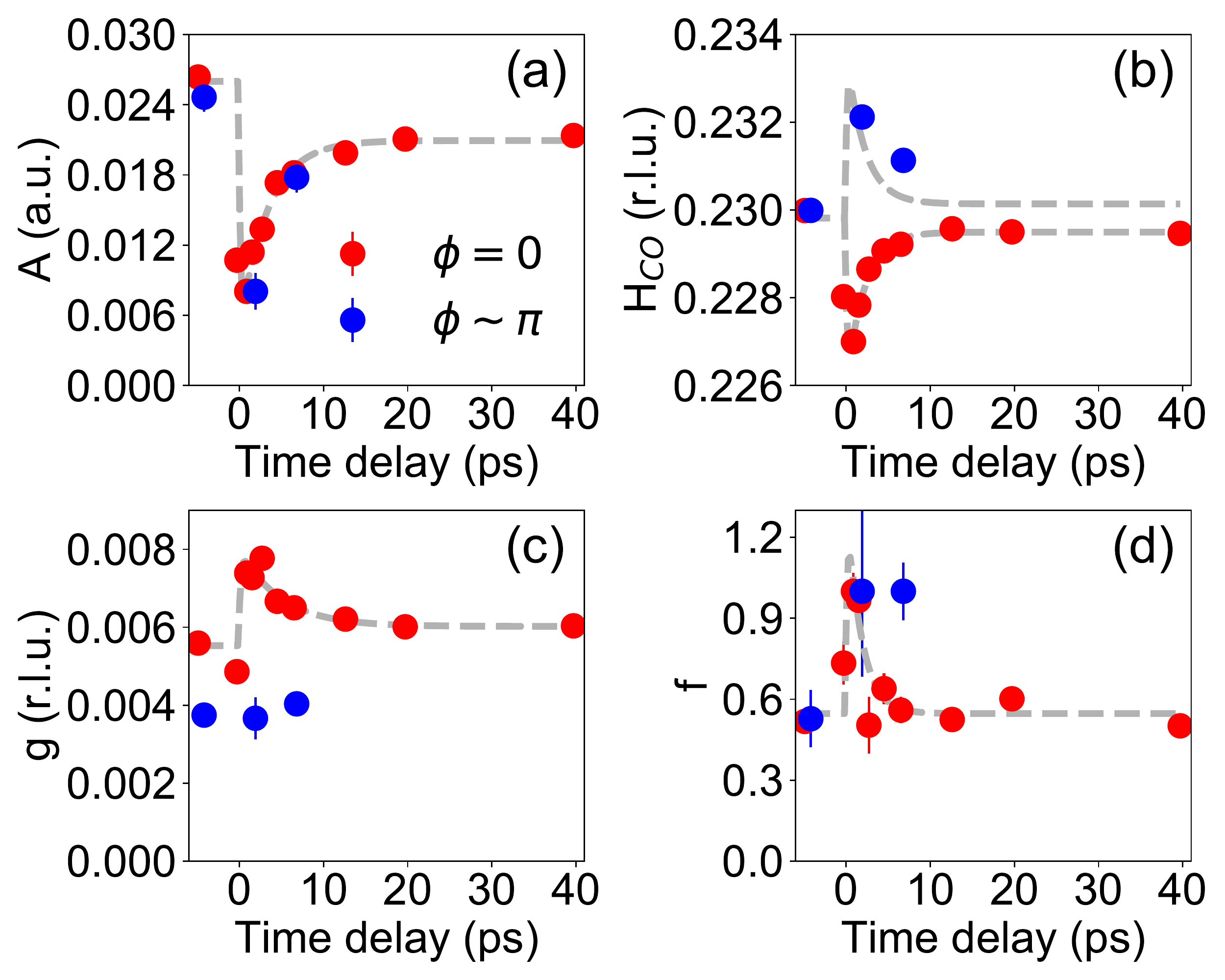}
    \caption{(color online) Time-dependent CO peak fit parameters at $\phi=0$ (red symbols) and $\phi\sim\pi$ (blue symbols). (a) Pseudo-Voigt amplitude A, (b) CO peak position along the $(H_{CO},0.00,1.50)$ direction, (c) CO peak HWHM g, and (d) Pseudo-Voigt mixing parameter f.}
    \label{fig:Recoil_parameters_PRB}
\end{figure}

We finally comment on the CO peak relaxation. The order parameter intensity equilibrates according to an exponential behavior [Fig. \ref{fig:Recoil_parameters_PRB}(a)] with a time constant $1/\gamma_0=(3.805\pm0.031)$ ps \cite{Mitrano2018}. However, a single exponential fit to the time-dependent momentum shift [Fig. \ref{fig:Recoil_parameters_PRB}(b)] yields a recovery time of $(2.13\pm0.18)$ ps, in line with the fast recovery observed in the quasielastic peak energy shift in Fig. \ref{fig:trRIXS_fit_parameters}(b). This implies that the pump-induced sliding does not follow the equilibration of the order parameter and cannot be captured by a conventional time-dependent Ginzburg-Landau description of the order parameter dynamics \cite{Mitrano2018}.

\section{Conclusion}
In summary, we report evidence of unconventional photoinduced dynamics in charge-ordered LBCO. By performing both energy-resolved and energy-integrated resonant soft x-ray scattering, we discovered the existence of a short-lived, nonequilibrium state, whose features are compatible with a sliding CDW coherently set in motion by the 1.55 eV pump pulses. This transient state is visible under mild excitation conditions (0.1 mJ/cm$^2$) and exhibits shifts in both the quasielastic line energy, and in the momentum wave vector. We crucially observe that the shift in momentum space tracks with the pump propagation direction, unlike previous experiments in conventional CDW systems. This feature is incompatible with a simple pump-induced deformation of the charge density modulation (e.g. by introducing a spatial phase gradient), which would not depend upon the pump orientation with respect to the sample. Therefore we interpret this phenomenon in terms of a Doppler shift due to the CO condensate moving with momentum $\Delta q$. The observation that ultrafast laser irradiation coherently sets the CO in motion like a classical rigid body is unprecedented, and provides a route to the coherent manipulation of electronic condensates at ultrafast timescales.

\section{Acknowledgments}
We acknowledge E. Fradkin, S. A. Kivelson, T. Devereaux, B. Moritz, H. Jang, S. Lee, J.-S. Lee, C. C. Kao, J. Turner, G. Dakovski, and Y. Y. Peng for valuable discussions, D. Swetz for help during the experiments, and S. Zohar for support in data analysis. This work was supported by U.S. Department of Energy, Office of Basic Energy Sciences grant no. DE-FG02-06ER46285. Use of the Linac Coherent Light Source (LCLS), SLAC National Accelerator Laboratory, was supported by DOE grant no. DE-AC02-76SF00515. Growth of LBCO single crystals was supported by DOE grant no. DE-SC0012704. M.M. acknowledges support from the Alexander von Humboldt foundation, and P.A. acknowledges support from the Gordon and Betty Moore Foundation's EPiQS initiative through grant GBMF-4542.

\appendix*
\section{Influence of the inelastic features on the quasielastic line shift}
\begin{figure}
    \centering
    \includegraphics[width=\columnwidth]{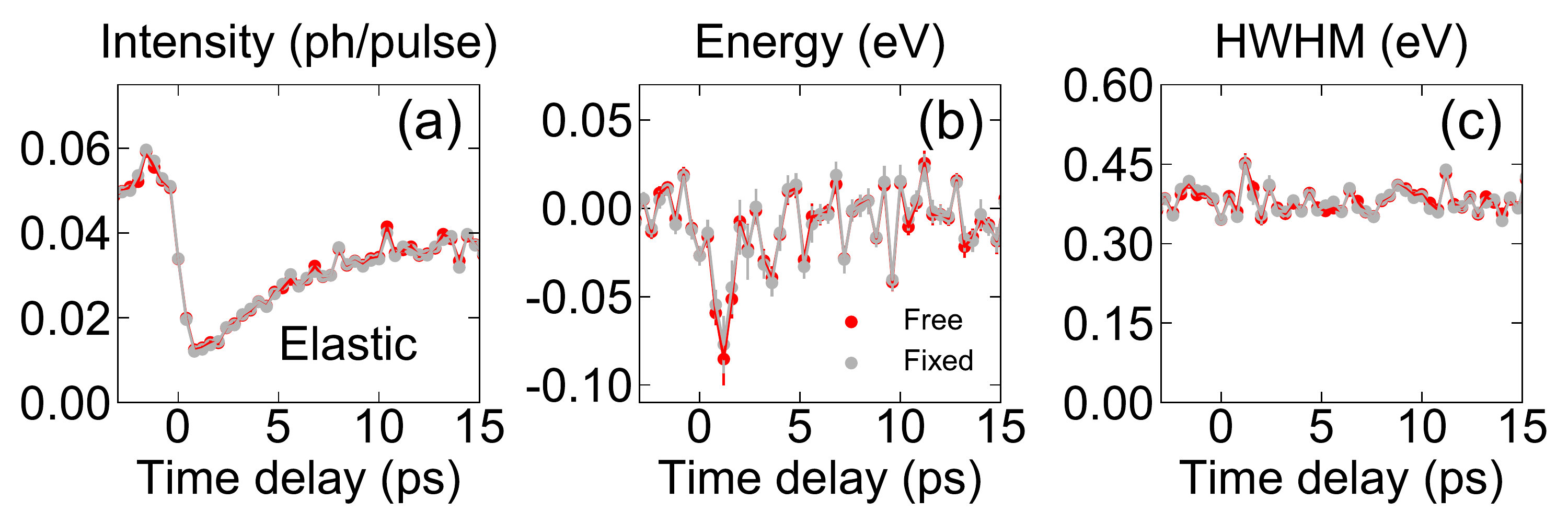}
    \caption{(color online) Fit parameters of the quasielastic line for free (red) and fixed (gray) parameters of both $dd/Cu^{2+}$ and $CT$ lines.}
    \label{fig:fit_comparison}
\end{figure}

In Figs. \ref{fig:trRIXS_map_fit} and \ref{fig:trRIXS_fit_parameters}, we report the observation of a 80-meV, pump-induced redshift of the quasielastic line and we characterize it with independent fits of the tr-RIXS spectra at each time delay. All fit parameters reported in Fig. \ref{fig:trRIXS_fit_parameters} are free to vary during the optimization cycles.\\
\begin{figure}
    \centering
    \includegraphics[width=\columnwidth]{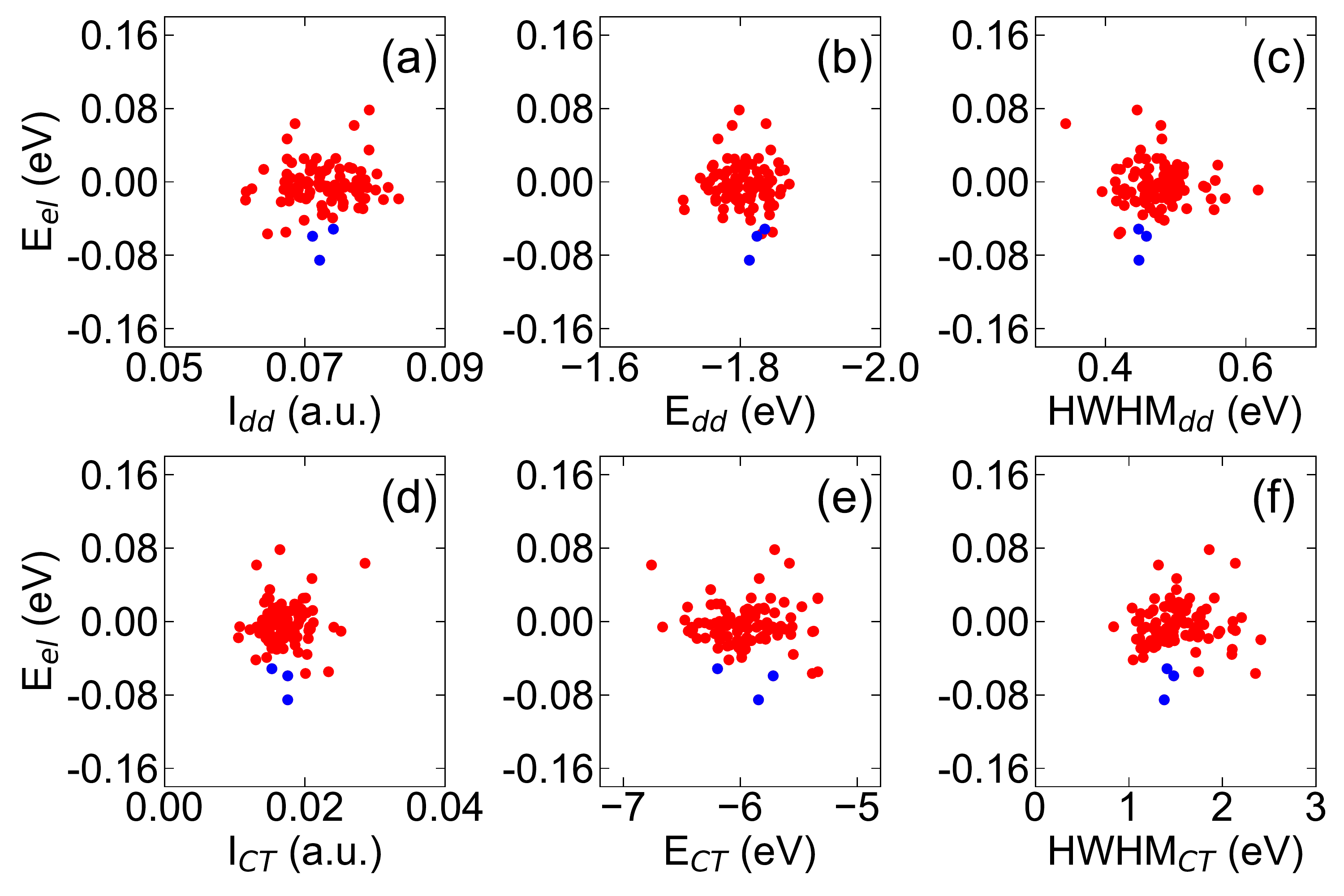}
    \caption{(color online) Correlation plots of the elastic line energy ($E_{el}$) with the intensity $I$, center energy  (E), and HWHM of the $dd/Cu^{2+}$ and $CT$ lines for all the sampled time delays (red symbols). Blue symbols identify the time delays corresponding to the maximum quasielastic energy shift.}
    \label{fig:correlation_plot}
\end{figure}
In order to exclude the possibility of artifacts due to the dynamics of the inelastic features and establish the shift as a genuine physical phenomenon, we verified the degree of correlation between the elastic line energy and the fit parameters for both $dd/Cu^{2+}$ and $CT$ lines (Fig. \ref{fig:correlation_plot}). All time delays cluster in a relatively symmetrical distribution, thus reflecting a low degree of correlation. This implies that the inelastic features have marginal, if any, effects on the quasielastic line dynamics. 
Furthermore, we repeated our fits with fixed inelastic parameters to evaluate the stability of our unconstrained fit procedure. When fixing the inelastic fit parameters to their mean values in Fig. \ref{fig:trRIXS_fit_parameters}(d)-\ref{fig:trRIXS_fit_parameters}(i), we obtain the same elastic line behavior and, crucially, the same 80-meV quasielastic redshift (see Fig. \ref{fig:fit_comparison}).
These tests confirm that the observed transient, quasielastic CO redshift is a true physical effect and it is not affected by the tails of the inelastic features of the LBCO RIXS spectrum.

\bibliography{LBCO-DynamicScaling_PRB_v5}

\begin{thebibliography}{39}%
\makeatletter
\providecommand \@ifxundefined [1]{%
 \@ifx{#1\undefined}
}%
\providecommand \@ifnum [1]{%
 \ifnum #1\expandafter \@firstoftwo
 \else \expandafter \@secondoftwo
 \fi
}%
\providecommand \@ifx [1]{%
 \ifx #1\expandafter \@firstoftwo
 \else \expandafter \@secondoftwo
 \fi
}%
\providecommand \natexlab [1]{#1}%
\providecommand \enquote  [1]{``#1''}%
\providecommand \bibnamefont  [1]{#1}%
\providecommand \bibfnamefont [1]{#1}%
\providecommand \citenamefont [1]{#1}%
\providecommand \href@noop [0]{\@secondoftwo}%
\providecommand \href [0]{\begingroup \@sanitize@url \@href}%
\providecommand \@href[1]{\@@startlink{#1}\@@href}%
\providecommand \@@href[1]{\endgroup#1\@@endlink}%
\providecommand \@sanitize@url [0]{\catcode `\\12\catcode `\$12\catcode
  `\&12\catcode `\#12\catcode `\^12\catcode `\_12\catcode `\%12\relax}%
\providecommand \@@startlink[1]{}%
\providecommand \@@endlink[0]{}%
\providecommand \url  [0]{\begingroup\@sanitize@url \@url }%
\providecommand \@url [1]{\endgroup\@href {#1}{\urlprefix }}%
\providecommand \urlprefix  [0]{URL }%
\providecommand \Eprint [0]{\href }%
\providecommand \doibase [0]{http://dx.doi.org/}%
\providecommand \selectlanguage [0]{\@gobble}%
\providecommand \bibinfo  [0]{\@secondoftwo}%
\providecommand \bibfield  [0]{\@secondoftwo}%
\providecommand \translation [1]{[#1]}%
\providecommand \BibitemOpen [0]{}%
\providecommand \bibitemStop [0]{}%
\providecommand \bibitemNoStop [0]{.\EOS\space}%
\providecommand \EOS [0]{\spacefactor3000\relax}%
\providecommand \BibitemShut  [1]{\csname bibitem#1\endcsname}%
\let\auto@bib@innerbib\@empty
\bibitem [{\citenamefont {Keimer}\ \emph {et~al.}(2015)\citenamefont {Keimer},
  \citenamefont {Kivelson}, \citenamefont {Norman}, \citenamefont {Uchida},\
  and\ \citenamefont {Zaanen}}]{Keimer2015}%
  \BibitemOpen
  \bibfield  {author} {\bibinfo {author} {\bibfnamefont {B.}~\bibnamefont
  {Keimer}}, \bibinfo {author} {\bibfnamefont {S.~A.}\ \bibnamefont
  {Kivelson}}, \bibinfo {author} {\bibfnamefont {M.~R.}\ \bibnamefont
  {Norman}}, \bibinfo {author} {\bibfnamefont {S.}~\bibnamefont {Uchida}}, \
  and\ \bibinfo {author} {\bibfnamefont {J.}~\bibnamefont {Zaanen}},\ }\href
  {\doibase 10.1038/nature14165} {\bibfield  {journal} {\bibinfo  {journal}
  {Nature}\ }\textbf {\bibinfo {volume} {518}},\ \bibinfo {pages} {179}
  (\bibinfo {year} {2015})}\BibitemShut {NoStop}%
\bibitem [{\citenamefont {Basov}\ \emph {et~al.}(2017)\citenamefont {Basov},
  \citenamefont {Averitt},\ and\ \citenamefont {Hsieh}}]{Basov2017}%
  \BibitemOpen
  \bibfield  {author} {\bibinfo {author} {\bibfnamefont {D.~N.}\ \bibnamefont
  {Basov}}, \bibinfo {author} {\bibfnamefont {R.~D.}\ \bibnamefont {Averitt}},
  \ and\ \bibinfo {author} {\bibfnamefont {D.}~\bibnamefont {Hsieh}},\ }\href
  {\doibase 10.1038/nmat5017} {\bibfield  {journal} {\bibinfo  {journal}
  {Nature Materials}\ }\textbf {\bibinfo {volume} {16}},\ \bibinfo {pages}
  {1077} (\bibinfo {year} {2017})}\BibitemShut {NoStop}%
\bibitem [{\citenamefont {Gr{\"u}ner}(1994)}]{Gruner}%
  \BibitemOpen
  \bibfield  {author} {\bibinfo {author} {\bibfnamefont {G.}~\bibnamefont
  {Gr{\"u}ner}},\ }\href@noop {} {\emph {\bibinfo {title} {Density Waves in
  Solids}}},\ Frontiers in Physics\ (\bibinfo  {publisher} {Perseus Books},\
  \bibinfo {address} {Cambridge, Massachussets, USA},\ \bibinfo {year}
  {1994})\BibitemShut {NoStop}%
\bibitem [{\citenamefont {Gr\"uner}(1988)}]{Gruner1988}%
  \BibitemOpen
  \bibfield  {author} {\bibinfo {author} {\bibfnamefont {G.}~\bibnamefont
  {Gr\"uner}},\ }\href {\doibase 10.1103/RevModPhys.60.1129} {\bibfield
  {journal} {\bibinfo  {journal} {Rev. Mod. Phys.}\ }\textbf {\bibinfo {volume}
  {60}},\ \bibinfo {pages} {1129} (\bibinfo {year} {1988})}\BibitemShut
  {NoStop}%
\bibitem [{\citenamefont {Monceau}(2012)}]{Monceau2012}%
  \BibitemOpen
  \bibfield  {author} {\bibinfo {author} {\bibfnamefont {P.}~\bibnamefont
  {Monceau}},\ }\href {\doibase 10.1080/00018732.2012.719674} {\bibfield
  {journal} {\bibinfo  {journal} {Advances in Physics}\ }\textbf {\bibinfo
  {volume} {61}},\ \bibinfo {pages} {325} (\bibinfo {year} {2012})}\BibitemShut
  {NoStop}%
\bibitem [{\citenamefont {DiCarlo}\ \emph {et~al.}(1993)\citenamefont
  {DiCarlo}, \citenamefont {Sweetland}, \citenamefont {Sutton}, \citenamefont
  {Brock},\ and\ \citenamefont {Thorne}}]{DiCarlo1993}%
  \BibitemOpen
  \bibfield  {author} {\bibinfo {author} {\bibfnamefont {D.}~\bibnamefont
  {DiCarlo}}, \bibinfo {author} {\bibfnamefont {E.}~\bibnamefont {Sweetland}},
  \bibinfo {author} {\bibfnamefont {M.}~\bibnamefont {Sutton}}, \bibinfo
  {author} {\bibfnamefont {J.~D.}\ \bibnamefont {Brock}}, \ and\ \bibinfo
  {author} {\bibfnamefont {R.~E.}\ \bibnamefont {Thorne}},\ }\href {\doibase
  10.1103/PhysRevLett.70.845} {\bibfield  {journal} {\bibinfo  {journal} {Phys.
  Rev. Lett.}\ }\textbf {\bibinfo {volume} {70}},\ \bibinfo {pages} {845}
  (\bibinfo {year} {1993})}\BibitemShut {NoStop}%
\bibitem [{\citenamefont {Requardt}\ \emph {et~al.}(1998)\citenamefont
  {Requardt}, \citenamefont {Nad}, \citenamefont {Monceau}, \citenamefont
  {Currat}, \citenamefont {Lorenzo}, \citenamefont {Brazovskii}, \citenamefont
  {Kirova}, \citenamefont {Gr\"ubel},\ and\ \citenamefont
  {Vettier}}]{Requardt1998}%
  \BibitemOpen
  \bibfield  {author} {\bibinfo {author} {\bibfnamefont {H.}~\bibnamefont
  {Requardt}}, \bibinfo {author} {\bibfnamefont {F.~Y.}\ \bibnamefont {Nad}},
  \bibinfo {author} {\bibfnamefont {P.}~\bibnamefont {Monceau}}, \bibinfo
  {author} {\bibfnamefont {R.}~\bibnamefont {Currat}}, \bibinfo {author}
  {\bibfnamefont {J.~E.}\ \bibnamefont {Lorenzo}}, \bibinfo {author}
  {\bibfnamefont {S.}~\bibnamefont {Brazovskii}}, \bibinfo {author}
  {\bibfnamefont {N.}~\bibnamefont {Kirova}}, \bibinfo {author} {\bibfnamefont
  {G.}~\bibnamefont {Gr\"ubel}}, \ and\ \bibinfo {author} {\bibfnamefont
  {C.}~\bibnamefont {Vettier}},\ }\href {\doibase 10.1103/PhysRevLett.80.5631}
  {\bibfield  {journal} {\bibinfo  {journal} {Phys. Rev. Lett.}\ }\textbf
  {\bibinfo {volume} {80}},\ \bibinfo {pages} {5631} (\bibinfo {year}
  {1998})}\BibitemShut {NoStop}%
\bibitem [{\citenamefont {Danneau}\ \emph {et~al.}(2002)\citenamefont
  {Danneau}, \citenamefont {Ayari}, \citenamefont {Rideau}, \citenamefont
  {Requardt}, \citenamefont {Lorenzo}, \citenamefont {Ortega}, \citenamefont
  {Monceau}, \citenamefont {Currat},\ and\ \citenamefont
  {Gr\"ubel.}}]{Danneau2002}%
  \BibitemOpen
  \bibfield  {author} {\bibinfo {author} {\bibfnamefont {R.}~\bibnamefont
  {Danneau}}, \bibinfo {author} {\bibfnamefont {A.}~\bibnamefont {Ayari}},
  \bibinfo {author} {\bibfnamefont {D.}~\bibnamefont {Rideau}}, \bibinfo
  {author} {\bibfnamefont {H.}~\bibnamefont {Requardt}}, \bibinfo {author}
  {\bibfnamefont {J.~E.}\ \bibnamefont {Lorenzo}}, \bibinfo {author}
  {\bibfnamefont {L.}~\bibnamefont {Ortega}}, \bibinfo {author} {\bibfnamefont
  {P.}~\bibnamefont {Monceau}}, \bibinfo {author} {\bibfnamefont
  {R.}~\bibnamefont {Currat}}, \ and\ \bibinfo {author} {\bibfnamefont
  {G.}~\bibnamefont {Gr\"ubel.}},\ }\href {\doibase
  10.1103/PhysRevLett.89.106404} {\bibfield  {journal} {\bibinfo  {journal}
  {Phys. Rev. Lett.}\ }\textbf {\bibinfo {volume} {89}},\ \bibinfo {pages}
  {106404} (\bibinfo {year} {2002})}\BibitemShut {NoStop}%
\bibitem [{\citenamefont {Pinsolle}\ \emph {et~al.}(2012)\citenamefont
  {Pinsolle}, \citenamefont {Kirova}, \citenamefont {Jacques}, \citenamefont
  {Sinchenko},\ and\ \citenamefont {Le~Bolloc'h}}]{Pinsolle2012}%
  \BibitemOpen
  \bibfield  {author} {\bibinfo {author} {\bibfnamefont {E.}~\bibnamefont
  {Pinsolle}}, \bibinfo {author} {\bibfnamefont {N.}~\bibnamefont {Kirova}},
  \bibinfo {author} {\bibfnamefont {V.~L.~R.}\ \bibnamefont {Jacques}},
  \bibinfo {author} {\bibfnamefont {A.~A.}\ \bibnamefont {Sinchenko}}, \ and\
  \bibinfo {author} {\bibfnamefont {D.}~\bibnamefont {Le~Bolloc'h}},\ }\href
  {\doibase 10.1103/PhysRevLett.109.256402} {\bibfield  {journal} {\bibinfo
  {journal} {Phys. Rev. Lett.}\ }\textbf {\bibinfo {volume} {109}},\ \bibinfo
  {pages} {256402} (\bibinfo {year} {2012})}\BibitemShut {NoStop}%
\bibitem [{\citenamefont {Tamegai}\ \emph {et~al.}(1984)\citenamefont
  {Tamegai}, \citenamefont {Tsutsumi}, \citenamefont {Kagoshima}, \citenamefont
  {Kanai}, \citenamefont {Tani}, \citenamefont {Tomozawa}, \citenamefont
  {Sato}, \citenamefont {Tsuji}, \citenamefont {Harada}, \citenamefont
  {Sakata},\ and\ \citenamefont {Nakajima}}]{Tamegai1984}%
  \BibitemOpen
  \bibfield  {author} {\bibinfo {author} {\bibfnamefont {T.}~\bibnamefont
  {Tamegai}}, \bibinfo {author} {\bibfnamefont {K.}~\bibnamefont {Tsutsumi}},
  \bibinfo {author} {\bibfnamefont {S.}~\bibnamefont {Kagoshima}}, \bibinfo
  {author} {\bibfnamefont {Y.}~\bibnamefont {Kanai}}, \bibinfo {author}
  {\bibfnamefont {M.}~\bibnamefont {Tani}}, \bibinfo {author} {\bibfnamefont
  {H.}~\bibnamefont {Tomozawa}}, \bibinfo {author} {\bibfnamefont
  {M.}~\bibnamefont {Sato}}, \bibinfo {author} {\bibfnamefont {K.}~\bibnamefont
  {Tsuji}}, \bibinfo {author} {\bibfnamefont {J.}~\bibnamefont {Harada}},
  \bibinfo {author} {\bibfnamefont {M.}~\bibnamefont {Sakata}}, \ and\ \bibinfo
  {author} {\bibfnamefont {T.}~\bibnamefont {Nakajima}},\ }\href {\doibase
  10.1016/0038-1098(84)91064-0} {\bibfield  {journal} {\bibinfo  {journal}
  {Solid State Communications}\ }\textbf {\bibinfo {volume} {51}},\ \bibinfo
  {pages} {585} (\bibinfo {year} {1984})}\BibitemShut {NoStop}%
\bibitem [{\citenamefont {Hundley}\ and\ \citenamefont
  {Zettl}(1989)}]{Hundley1989}%
  \BibitemOpen
  \bibfield  {author} {\bibinfo {author} {\bibfnamefont {M.~F.}\ \bibnamefont
  {Hundley}}\ and\ \bibinfo {author} {\bibfnamefont {A.}~\bibnamefont
  {Zettl}},\ }\href {\doibase 10.1103/PhysRevB.39.3026} {\bibfield  {journal}
  {\bibinfo  {journal} {Phys. Rev. B}\ }\textbf {\bibinfo {volume} {39}},\
  \bibinfo {pages} {3026} (\bibinfo {year} {1989})}\BibitemShut {NoStop}%
\bibitem [{\citenamefont {Jacques}\ \emph {et~al.}(2016)\citenamefont
  {Jacques}, \citenamefont {Laulh\'e}, \citenamefont {Moisan}, \citenamefont
  {Ravy},\ and\ \citenamefont {Le~Bolloc'h}}]{Jacques2016}%
  \BibitemOpen
  \bibfield  {author} {\bibinfo {author} {\bibfnamefont {V.~L.~R.}\
  \bibnamefont {Jacques}}, \bibinfo {author} {\bibfnamefont {C.}~\bibnamefont
  {Laulh\'e}}, \bibinfo {author} {\bibfnamefont {N.}~\bibnamefont {Moisan}},
  \bibinfo {author} {\bibfnamefont {S.}~\bibnamefont {Ravy}}, \ and\ \bibinfo
  {author} {\bibfnamefont {D.}~\bibnamefont {Le~Bolloc'h}},\ }\href {\doibase
  10.1103/PhysRevLett.117.156401} {\bibfield  {journal} {\bibinfo  {journal}
  {Phys. Rev. Lett.}\ }\textbf {\bibinfo {volume} {117}},\ \bibinfo {pages}
  {156401} (\bibinfo {year} {2016})}\BibitemShut {NoStop}%
\bibitem [{\citenamefont {Torchinsky}\ \emph {et~al.}(2013)\citenamefont
  {Torchinsky}, \citenamefont {Mahmood}, \citenamefont {Bollinger},
  \citenamefont {Bo{\v z}ovi{\'c}},\ and\ \citenamefont
  {Gedik}}]{Torchinsky2013}%
  \BibitemOpen
  \bibfield  {author} {\bibinfo {author} {\bibfnamefont {D.~H.}\ \bibnamefont
  {Torchinsky}}, \bibinfo {author} {\bibfnamefont {F.}~\bibnamefont {Mahmood}},
  \bibinfo {author} {\bibfnamefont {A.~T.}\ \bibnamefont {Bollinger}}, \bibinfo
  {author} {\bibfnamefont {I.}~\bibnamefont {Bo{\v z}ovi{\'c}}}, \ and\
  \bibinfo {author} {\bibfnamefont {N.}~\bibnamefont {Gedik}},\ }\href
  {\doibase 10.1038/nmat3571} {\bibfield  {journal} {\bibinfo  {journal}
  {Nature Materials}\ }\textbf {\bibinfo {volume} {12}},\ \bibinfo {pages}
  {387} (\bibinfo {year} {2013})}\BibitemShut {NoStop}%
\bibitem [{\citenamefont {Dakovski}\ \emph {et~al.}(2015)\citenamefont
  {Dakovski}, \citenamefont {Lee}, \citenamefont {Hawthorn}, \citenamefont
  {Garner}, \citenamefont {Bonn}, \citenamefont {Hardy}, \citenamefont {Liang},
  \citenamefont {Hoffmann},\ and\ \citenamefont {Turner}}]{Dakovski2015}%
  \BibitemOpen
  \bibfield  {author} {\bibinfo {author} {\bibfnamefont {G.~L.}\ \bibnamefont
  {Dakovski}}, \bibinfo {author} {\bibfnamefont {W.-S.}\ \bibnamefont {Lee}},
  \bibinfo {author} {\bibfnamefont {D.~G.}\ \bibnamefont {Hawthorn}}, \bibinfo
  {author} {\bibfnamefont {N.}~\bibnamefont {Garner}}, \bibinfo {author}
  {\bibfnamefont {D.}~\bibnamefont {Bonn}}, \bibinfo {author} {\bibfnamefont
  {W.}~\bibnamefont {Hardy}}, \bibinfo {author} {\bibfnamefont
  {R.}~\bibnamefont {Liang}}, \bibinfo {author} {\bibfnamefont {M.~C.}\
  \bibnamefont {Hoffmann}}, \ and\ \bibinfo {author} {\bibfnamefont {J.~J.}\
  \bibnamefont {Turner}},\ }\href {\doibase 10.1103/PhysRevB.91.220506}
  {\bibfield  {journal} {\bibinfo  {journal} {Phys. Rev. B}\ }\textbf {\bibinfo
  {volume} {91}},\ \bibinfo {pages} {220506} (\bibinfo {year}
  {2015})}\BibitemShut {NoStop}%
\bibitem [{\citenamefont {Hinton}\ \emph {et~al.}(2013)\citenamefont {Hinton},
  \citenamefont {Koralek}, \citenamefont {Lu}, \citenamefont {Vishwanath},
  \citenamefont {Orenstein}, \citenamefont {Bonn}, \citenamefont {Hardy},\ and\
  \citenamefont {Liang}}]{Hinton2013}%
  \BibitemOpen
  \bibfield  {author} {\bibinfo {author} {\bibfnamefont {J.~P.}\ \bibnamefont
  {Hinton}}, \bibinfo {author} {\bibfnamefont {J.~D.}\ \bibnamefont {Koralek}},
  \bibinfo {author} {\bibfnamefont {Y.~M.}\ \bibnamefont {Lu}}, \bibinfo
  {author} {\bibfnamefont {A.}~\bibnamefont {Vishwanath}}, \bibinfo {author}
  {\bibfnamefont {J.}~\bibnamefont {Orenstein}}, \bibinfo {author}
  {\bibfnamefont {D.~A.}\ \bibnamefont {Bonn}}, \bibinfo {author}
  {\bibfnamefont {W.~N.}\ \bibnamefont {Hardy}}, \ and\ \bibinfo {author}
  {\bibfnamefont {R.}~\bibnamefont {Liang}},\ }\href {\doibase
  10.1103/PhysRevB.88.060508} {\bibfield  {journal} {\bibinfo  {journal} {Phys.
  Rev. B}\ }\textbf {\bibinfo {volume} {88}},\ \bibinfo {pages} {060508}
  (\bibinfo {year} {2013})}\BibitemShut {NoStop}%
\bibitem [{\citenamefont {Huber}\ \emph {et~al.}(2014)\citenamefont {Huber},
  \citenamefont {Mariager}, \citenamefont {Ferrer}, \citenamefont {Sch\"afer},
  \citenamefont {Johnson}, \citenamefont {Gr\"ubel}, \citenamefont {L\"ubcke},
  \citenamefont {Huber}, \citenamefont {Kubacka}, \citenamefont {Dornes},
  \citenamefont {Laulhe}, \citenamefont {Ravy}, \citenamefont {Ingold},
  \citenamefont {Beaud}, \citenamefont {Demsar},\ and\ \citenamefont
  {Johnson}}]{Huber2014}%
  \BibitemOpen
  \bibfield  {author} {\bibinfo {author} {\bibfnamefont {T.}~\bibnamefont
  {Huber}}, \bibinfo {author} {\bibfnamefont {S.~O.}\ \bibnamefont {Mariager}},
  \bibinfo {author} {\bibfnamefont {A.}~\bibnamefont {Ferrer}}, \bibinfo
  {author} {\bibfnamefont {H.}~\bibnamefont {Sch\"afer}}, \bibinfo {author}
  {\bibfnamefont {J.~A.}\ \bibnamefont {Johnson}}, \bibinfo {author}
  {\bibfnamefont {S.}~\bibnamefont {Gr\"ubel}}, \bibinfo {author}
  {\bibfnamefont {A.}~\bibnamefont {L\"ubcke}}, \bibinfo {author}
  {\bibfnamefont {L.}~\bibnamefont {Huber}}, \bibinfo {author} {\bibfnamefont
  {T.}~\bibnamefont {Kubacka}}, \bibinfo {author} {\bibfnamefont
  {C.}~\bibnamefont {Dornes}}, \bibinfo {author} {\bibfnamefont
  {C.}~\bibnamefont {Laulhe}}, \bibinfo {author} {\bibfnamefont
  {S.}~\bibnamefont {Ravy}}, \bibinfo {author} {\bibfnamefont {G.}~\bibnamefont
  {Ingold}}, \bibinfo {author} {\bibfnamefont {P.}~\bibnamefont {Beaud}},
  \bibinfo {author} {\bibfnamefont {J.}~\bibnamefont {Demsar}}, \ and\ \bibinfo
  {author} {\bibfnamefont {S.~L.}\ \bibnamefont {Johnson}},\ }\href {\doibase
  10.1103/PhysRevLett.113.026401} {\bibfield  {journal} {\bibinfo  {journal}
  {Phys. Rev. Lett.}\ }\textbf {\bibinfo {volume} {113}},\ \bibinfo {pages}
  {026401} (\bibinfo {year} {2014})}\BibitemShut {NoStop}%
\bibitem [{\citenamefont {Beaud}\ \emph {et~al.}(2014)\citenamefont {Beaud},
  \citenamefont {Caviezel}, \citenamefont {Mariager}, \citenamefont {Rettig},
  \citenamefont {Ingold}, \citenamefont {Dornes}, \citenamefont {Huang},
  \citenamefont {Johnson}, \citenamefont {Radovic}, \citenamefont {Huber},
  \citenamefont {Kubacka}, \citenamefont {Ferrer}, \citenamefont {Lemke},
  \citenamefont {Chollet}, \citenamefont {Zhu}, \citenamefont {Glownia},
  \citenamefont {Sikorski}, \citenamefont {Robert}, \citenamefont {Wadati},
  \citenamefont {Nakamura}, \citenamefont {Kawasaki}, \citenamefont {Tokura},
  \citenamefont {Johnson},\ and\ \citenamefont {Staub}}]{Beaud2014}%
  \BibitemOpen
  \bibfield  {author} {\bibinfo {author} {\bibfnamefont {P.}~\bibnamefont
  {Beaud}}, \bibinfo {author} {\bibfnamefont {A.}~\bibnamefont {Caviezel}},
  \bibinfo {author} {\bibfnamefont {S.~O.}\ \bibnamefont {Mariager}}, \bibinfo
  {author} {\bibfnamefont {L.}~\bibnamefont {Rettig}}, \bibinfo {author}
  {\bibfnamefont {G.}~\bibnamefont {Ingold}}, \bibinfo {author} {\bibfnamefont
  {C.}~\bibnamefont {Dornes}}, \bibinfo {author} {\bibfnamefont {S.-W.}\
  \bibnamefont {Huang}}, \bibinfo {author} {\bibfnamefont {J.~A.}\ \bibnamefont
  {Johnson}}, \bibinfo {author} {\bibfnamefont {M.}~\bibnamefont {Radovic}},
  \bibinfo {author} {\bibfnamefont {T.}~\bibnamefont {Huber}}, \bibinfo
  {author} {\bibfnamefont {T.}~\bibnamefont {Kubacka}}, \bibinfo {author}
  {\bibfnamefont {A.}~\bibnamefont {Ferrer}}, \bibinfo {author} {\bibfnamefont
  {H.~T.}\ \bibnamefont {Lemke}}, \bibinfo {author} {\bibfnamefont
  {M.}~\bibnamefont {Chollet}}, \bibinfo {author} {\bibfnamefont
  {D.}~\bibnamefont {Zhu}}, \bibinfo {author} {\bibfnamefont {J.~M.}\
  \bibnamefont {Glownia}}, \bibinfo {author} {\bibfnamefont {M.}~\bibnamefont
  {Sikorski}}, \bibinfo {author} {\bibfnamefont {A.}~\bibnamefont {Robert}},
  \bibinfo {author} {\bibfnamefont {H.}~\bibnamefont {Wadati}}, \bibinfo
  {author} {\bibfnamefont {M.}~\bibnamefont {Nakamura}}, \bibinfo {author}
  {\bibfnamefont {M.}~\bibnamefont {Kawasaki}}, \bibinfo {author}
  {\bibfnamefont {Y.}~\bibnamefont {Tokura}}, \bibinfo {author} {\bibfnamefont
  {S.~L.}\ \bibnamefont {Johnson}}, \ and\ \bibinfo {author} {\bibfnamefont
  {U.}~\bibnamefont {Staub}},\ }\href {\doibase 10.1038/nmat4046} {\bibfield
  {journal} {\bibinfo  {journal} {Nat. Mater.}\ }\textbf {\bibinfo {volume}
  {13}},\ \bibinfo {pages} {923} (\bibinfo {year} {2014})}\BibitemShut
  {NoStop}%
\bibitem [{\citenamefont {Lee}\ \emph {et~al.}(2012)\citenamefont {Lee},
  \citenamefont {Chuang}, \citenamefont {Moore}, \citenamefont {Zhu},
  \citenamefont {Patthey}, \citenamefont {Trigo}, \citenamefont {Lu},
  \citenamefont {Kirchmann}, \citenamefont {Krupin}, \citenamefont {Yi},
  \citenamefont {Langner}, \citenamefont {Huse}, \citenamefont {Robinson},
  \citenamefont {Chen}, \citenamefont {Zhou}, \citenamefont {Coslovich},
  \citenamefont {Huber}, \citenamefont {Reis}, \citenamefont {Kaindl},
  \citenamefont {Schoenlein}, \citenamefont {Doering}, \citenamefont {Denes},
  \citenamefont {Schlotter}, \citenamefont {Turner}, \citenamefont {Johnson},
  \citenamefont {F{\"o}rst}, \citenamefont {Sasagawa}, \citenamefont {Kung},
  \citenamefont {Sorini}, \citenamefont {Kemper}, \citenamefont {Moritz},
  \citenamefont {Devereaux}, \citenamefont {Lee}, \citenamefont {Shen},\ and\
  \citenamefont {Hussain}}]{Lee2012}%
  \BibitemOpen
  \bibfield  {author} {\bibinfo {author} {\bibfnamefont {W.~S.}\ \bibnamefont
  {Lee}}, \bibinfo {author} {\bibfnamefont {Y.~D.}\ \bibnamefont {Chuang}},
  \bibinfo {author} {\bibfnamefont {R.~G.}\ \bibnamefont {Moore}}, \bibinfo
  {author} {\bibfnamefont {Y.}~\bibnamefont {Zhu}}, \bibinfo {author}
  {\bibfnamefont {L.}~\bibnamefont {Patthey}}, \bibinfo {author} {\bibfnamefont
  {M.}~\bibnamefont {Trigo}}, \bibinfo {author} {\bibfnamefont {D.~H.}\
  \bibnamefont {Lu}}, \bibinfo {author} {\bibfnamefont {P.~S.}\ \bibnamefont
  {Kirchmann}}, \bibinfo {author} {\bibfnamefont {O.}~\bibnamefont {Krupin}},
  \bibinfo {author} {\bibfnamefont {M.}~\bibnamefont {Yi}}, \bibinfo {author}
  {\bibfnamefont {M.}~\bibnamefont {Langner}}, \bibinfo {author} {\bibfnamefont
  {N.}~\bibnamefont {Huse}}, \bibinfo {author} {\bibfnamefont {J.~S.}\
  \bibnamefont {Robinson}}, \bibinfo {author} {\bibfnamefont {Y.}~\bibnamefont
  {Chen}}, \bibinfo {author} {\bibfnamefont {S.~Y.}\ \bibnamefont {Zhou}},
  \bibinfo {author} {\bibfnamefont {G.}~\bibnamefont {Coslovich}}, \bibinfo
  {author} {\bibfnamefont {B.}~\bibnamefont {Huber}}, \bibinfo {author}
  {\bibfnamefont {D.~A.}\ \bibnamefont {Reis}}, \bibinfo {author}
  {\bibfnamefont {R.~A.}\ \bibnamefont {Kaindl}}, \bibinfo {author}
  {\bibfnamefont {R.~W.}\ \bibnamefont {Schoenlein}}, \bibinfo {author}
  {\bibfnamefont {D.}~\bibnamefont {Doering}}, \bibinfo {author} {\bibfnamefont
  {P.}~\bibnamefont {Denes}}, \bibinfo {author} {\bibfnamefont {W.~F.}\
  \bibnamefont {Schlotter}}, \bibinfo {author} {\bibfnamefont {J.~J.}\
  \bibnamefont {Turner}}, \bibinfo {author} {\bibfnamefont {S.~L.}\
  \bibnamefont {Johnson}}, \bibinfo {author} {\bibfnamefont {M.}~\bibnamefont
  {F{\"o}rst}}, \bibinfo {author} {\bibfnamefont {T.}~\bibnamefont {Sasagawa}},
  \bibinfo {author} {\bibfnamefont {Y.~F.}\ \bibnamefont {Kung}}, \bibinfo
  {author} {\bibfnamefont {A.~P.}\ \bibnamefont {Sorini}}, \bibinfo {author}
  {\bibfnamefont {A.~F.}\ \bibnamefont {Kemper}}, \bibinfo {author}
  {\bibfnamefont {B.}~\bibnamefont {Moritz}}, \bibinfo {author} {\bibfnamefont
  {T.~P.}\ \bibnamefont {Devereaux}}, \bibinfo {author} {\bibfnamefont {D.~H.}\
  \bibnamefont {Lee}}, \bibinfo {author} {\bibfnamefont {Z.-X.}\ \bibnamefont
  {Shen}}, \ and\ \bibinfo {author} {\bibfnamefont {Z.}~\bibnamefont
  {Hussain}},\ }\href {\doibase 10.1038/ncomms1837} {\bibfield  {journal}
  {\bibinfo  {journal} {Nature Communications}\ }\textbf {\bibinfo {volume}
  {3}},\ \bibinfo {pages} {838} (\bibinfo {year} {2012})}\BibitemShut {NoStop}%
\bibitem [{\citenamefont {Chuang}\ \emph {et~al.}(2013)\citenamefont {Chuang},
  \citenamefont {Lee}, \citenamefont {Kung}, \citenamefont {Sorini},
  \citenamefont {Moritz}, \citenamefont {Moore}, \citenamefont {Patthey},
  \citenamefont {Trigo}, \citenamefont {Lu}, \citenamefont {Kirchmann},
  \citenamefont {Yi}, \citenamefont {Krupin}, \citenamefont {Langner},
  \citenamefont {Zhu}, \citenamefont {Zhou}, \citenamefont {Reis},
  \citenamefont {Huse}, \citenamefont {Robinson}, \citenamefont {Kaindl},
  \citenamefont {Schoenlein}, \citenamefont {Johnson}, \citenamefont {F\"orst},
  \citenamefont {Doering}, \citenamefont {Denes}, \citenamefont {Schlotter},
  \citenamefont {Turner}, \citenamefont {Sasagawa}, \citenamefont {Hussain},
  \citenamefont {Shen},\ and\ \citenamefont {Devereaux}}]{Chuang2013}%
  \BibitemOpen
  \bibfield  {author} {\bibinfo {author} {\bibfnamefont {Y.~D.}\ \bibnamefont
  {Chuang}}, \bibinfo {author} {\bibfnamefont {W.~S.}\ \bibnamefont {Lee}},
  \bibinfo {author} {\bibfnamefont {Y.~F.}\ \bibnamefont {Kung}}, \bibinfo
  {author} {\bibfnamefont {A.~P.}\ \bibnamefont {Sorini}}, \bibinfo {author}
  {\bibfnamefont {B.}~\bibnamefont {Moritz}}, \bibinfo {author} {\bibfnamefont
  {R.~G.}\ \bibnamefont {Moore}}, \bibinfo {author} {\bibfnamefont
  {L.}~\bibnamefont {Patthey}}, \bibinfo {author} {\bibfnamefont
  {M.}~\bibnamefont {Trigo}}, \bibinfo {author} {\bibfnamefont {D.~H.}\
  \bibnamefont {Lu}}, \bibinfo {author} {\bibfnamefont {P.~S.}\ \bibnamefont
  {Kirchmann}}, \bibinfo {author} {\bibfnamefont {M.}~\bibnamefont {Yi}},
  \bibinfo {author} {\bibfnamefont {O.}~\bibnamefont {Krupin}}, \bibinfo
  {author} {\bibfnamefont {M.}~\bibnamefont {Langner}}, \bibinfo {author}
  {\bibfnamefont {Y.}~\bibnamefont {Zhu}}, \bibinfo {author} {\bibfnamefont
  {S.~Y.}\ \bibnamefont {Zhou}}, \bibinfo {author} {\bibfnamefont {D.~A.}\
  \bibnamefont {Reis}}, \bibinfo {author} {\bibfnamefont {N.}~\bibnamefont
  {Huse}}, \bibinfo {author} {\bibfnamefont {J.~S.}\ \bibnamefont {Robinson}},
  \bibinfo {author} {\bibfnamefont {R.~A.}\ \bibnamefont {Kaindl}}, \bibinfo
  {author} {\bibfnamefont {R.~W.}\ \bibnamefont {Schoenlein}}, \bibinfo
  {author} {\bibfnamefont {S.~L.}\ \bibnamefont {Johnson}}, \bibinfo {author}
  {\bibfnamefont {M.}~\bibnamefont {F\"orst}}, \bibinfo {author} {\bibfnamefont
  {D.}~\bibnamefont {Doering}}, \bibinfo {author} {\bibfnamefont
  {P.}~\bibnamefont {Denes}}, \bibinfo {author} {\bibfnamefont {W.~F.}\
  \bibnamefont {Schlotter}}, \bibinfo {author} {\bibfnamefont {J.~J.}\
  \bibnamefont {Turner}}, \bibinfo {author} {\bibfnamefont {T.}~\bibnamefont
  {Sasagawa}}, \bibinfo {author} {\bibfnamefont {Z.}~\bibnamefont {Hussain}},
  \bibinfo {author} {\bibfnamefont {Z.~X.}\ \bibnamefont {Shen}}, \ and\
  \bibinfo {author} {\bibfnamefont {T.~P.}\ \bibnamefont {Devereaux}},\ }\href
  {\doibase 10.1103/PhysRevLett.110.127404} {\bibfield  {journal} {\bibinfo
  {journal} {Phys. Rev. Lett.}\ }\textbf {\bibinfo {volume} {110}},\ \bibinfo
  {pages} {127404} (\bibinfo {year} {2013})}\BibitemShut {NoStop}%
\bibitem [{\citenamefont {Fausti}\ \emph {et~al.}(2011)\citenamefont {Fausti},
  \citenamefont {Tobey}, \citenamefont {Dean}, \citenamefont {Kaiser},
  \citenamefont {Dienst}, \citenamefont {Hoffmann}, \citenamefont {Pyon},
  \citenamefont {Takayama}, \citenamefont {Takagi},\ and\ \citenamefont
  {Cavalleri}}]{Fausti2011}%
  \BibitemOpen
  \bibfield  {author} {\bibinfo {author} {\bibfnamefont {D.}~\bibnamefont
  {Fausti}}, \bibinfo {author} {\bibfnamefont {R.~I.}\ \bibnamefont {Tobey}},
  \bibinfo {author} {\bibfnamefont {N.}~\bibnamefont {Dean}}, \bibinfo {author}
  {\bibfnamefont {S.}~\bibnamefont {Kaiser}}, \bibinfo {author} {\bibfnamefont
  {A.}~\bibnamefont {Dienst}}, \bibinfo {author} {\bibfnamefont {M.~C.}\
  \bibnamefont {Hoffmann}}, \bibinfo {author} {\bibfnamefont {S.}~\bibnamefont
  {Pyon}}, \bibinfo {author} {\bibfnamefont {T.}~\bibnamefont {Takayama}},
  \bibinfo {author} {\bibfnamefont {H.}~\bibnamefont {Takagi}}, \ and\ \bibinfo
  {author} {\bibfnamefont {A.}~\bibnamefont {Cavalleri}},\ }\href {\doibase
  10.1126/science.1197294} {\bibfield  {journal} {\bibinfo  {journal}
  {Science}\ }\textbf {\bibinfo {volume} {331}},\ \bibinfo {pages} {189}
  (\bibinfo {year} {2011})}\BibitemShut {NoStop}%
\bibitem [{\citenamefont {Hu}\ \emph {et~al.}(2014)\citenamefont {Hu},
  \citenamefont {Kaiser}, \citenamefont {Nicoletti}, \citenamefont {Hunt},
  \citenamefont {Gierz}, \citenamefont {Hoffmann}, \citenamefont {Le~Tacon},
  \citenamefont {Loew}, \citenamefont {Keimer},\ and\ \citenamefont
  {Cavalleri}}]{Hu2014}%
  \BibitemOpen
  \bibfield  {author} {\bibinfo {author} {\bibfnamefont {W.}~\bibnamefont
  {Hu}}, \bibinfo {author} {\bibfnamefont {S.}~\bibnamefont {Kaiser}}, \bibinfo
  {author} {\bibfnamefont {D.}~\bibnamefont {Nicoletti}}, \bibinfo {author}
  {\bibfnamefont {C.~R.}\ \bibnamefont {Hunt}}, \bibinfo {author}
  {\bibfnamefont {I.}~\bibnamefont {Gierz}}, \bibinfo {author} {\bibfnamefont
  {M.~C.}\ \bibnamefont {Hoffmann}}, \bibinfo {author} {\bibfnamefont
  {M.}~\bibnamefont {Le~Tacon}}, \bibinfo {author} {\bibfnamefont
  {T.}~\bibnamefont {Loew}}, \bibinfo {author} {\bibfnamefont {B.}~\bibnamefont
  {Keimer}}, \ and\ \bibinfo {author} {\bibfnamefont {A.}~\bibnamefont
  {Cavalleri}},\ }\href {\doibase 10.1038/nmat3963} {\bibfield  {journal}
  {\bibinfo  {journal} {Nature Materials}\ }\textbf {\bibinfo {volume} {13}},\
  \bibinfo {pages} {705} (\bibinfo {year} {2014})}\BibitemShut {NoStop}%
\bibitem [{\citenamefont {Kaiser}\ \emph {et~al.}(2014)\citenamefont {Kaiser},
  \citenamefont {Hunt}, \citenamefont {Nicoletti}, \citenamefont {Hu},
  \citenamefont {Gierz}, \citenamefont {Liu}, \citenamefont {Le~Tacon},
  \citenamefont {Loew}, \citenamefont {Haug}, \citenamefont {Keimer},\ and\
  \citenamefont {Cavalleri}}]{Kaiser2014}%
  \BibitemOpen
  \bibfield  {author} {\bibinfo {author} {\bibfnamefont {S.}~\bibnamefont
  {Kaiser}}, \bibinfo {author} {\bibfnamefont {C.~R.}\ \bibnamefont {Hunt}},
  \bibinfo {author} {\bibfnamefont {D.}~\bibnamefont {Nicoletti}}, \bibinfo
  {author} {\bibfnamefont {W.}~\bibnamefont {Hu}}, \bibinfo {author}
  {\bibfnamefont {I.}~\bibnamefont {Gierz}}, \bibinfo {author} {\bibfnamefont
  {H.~Y.}\ \bibnamefont {Liu}}, \bibinfo {author} {\bibfnamefont
  {M.}~\bibnamefont {Le~Tacon}}, \bibinfo {author} {\bibfnamefont
  {T.}~\bibnamefont {Loew}}, \bibinfo {author} {\bibfnamefont {D.}~\bibnamefont
  {Haug}}, \bibinfo {author} {\bibfnamefont {B.}~\bibnamefont {Keimer}}, \ and\
  \bibinfo {author} {\bibfnamefont {A.}~\bibnamefont {Cavalleri}},\ }\href
  {\doibase 10.1103/PhysRevB.89.184516} {\bibfield  {journal} {\bibinfo
  {journal} {Phys. Rev. B}\ }\textbf {\bibinfo {volume} {89}},\ \bibinfo
  {pages} {184516} (\bibinfo {year} {2014})}\BibitemShut {NoStop}%
\bibitem [{\citenamefont {Nicoletti}\ \emph {et~al.}(2014)\citenamefont
  {Nicoletti}, \citenamefont {Casandruc}, \citenamefont {Laplace},
  \citenamefont {Khanna}, \citenamefont {Hunt}, \citenamefont {Kaiser},
  \citenamefont {Dhesi}, \citenamefont {Gu}, \citenamefont {Hill},\ and\
  \citenamefont {Cavalleri}}]{Nicoletti2014}%
  \BibitemOpen
  \bibfield  {author} {\bibinfo {author} {\bibfnamefont {D.}~\bibnamefont
  {Nicoletti}}, \bibinfo {author} {\bibfnamefont {E.}~\bibnamefont
  {Casandruc}}, \bibinfo {author} {\bibfnamefont {Y.}~\bibnamefont {Laplace}},
  \bibinfo {author} {\bibfnamefont {V.}~\bibnamefont {Khanna}}, \bibinfo
  {author} {\bibfnamefont {C.~R.}\ \bibnamefont {Hunt}}, \bibinfo {author}
  {\bibfnamefont {S.}~\bibnamefont {Kaiser}}, \bibinfo {author} {\bibfnamefont
  {S.~S.}\ \bibnamefont {Dhesi}}, \bibinfo {author} {\bibfnamefont {G.~D.}\
  \bibnamefont {Gu}}, \bibinfo {author} {\bibfnamefont {J.~P.}\ \bibnamefont
  {Hill}}, \ and\ \bibinfo {author} {\bibfnamefont {A.}~\bibnamefont
  {Cavalleri}},\ }\href {\doibase 10.1103/PhysRevB.90.100503} {\bibfield
  {journal} {\bibinfo  {journal} {Phys. Rev. B}\ }\textbf {\bibinfo {volume}
  {90}},\ \bibinfo {pages} {100503} (\bibinfo {year} {2014})}\BibitemShut
  {NoStop}%
\bibitem [{\citenamefont {F\"orst}\ \emph {et~al.}(2014)\citenamefont
  {F\"orst}, \citenamefont {Tobey}, \citenamefont {Bromberger}, \citenamefont
  {Wilkins}, \citenamefont {Khanna}, \citenamefont {Caviglia}, \citenamefont
  {Chuang}, \citenamefont {Lee}, \citenamefont {Schlotter}, \citenamefont
  {Turner}, \citenamefont {Minitti}, \citenamefont {Krupin}, \citenamefont
  {Xu}, \citenamefont {Wen}, \citenamefont {Gu}, \citenamefont {Dhesi},
  \citenamefont {Cavalleri},\ and\ \citenamefont {Hill}}]{Forst2014}%
  \BibitemOpen
  \bibfield  {author} {\bibinfo {author} {\bibfnamefont {M.}~\bibnamefont
  {F\"orst}}, \bibinfo {author} {\bibfnamefont {R.~I.}\ \bibnamefont {Tobey}},
  \bibinfo {author} {\bibfnamefont {H.}~\bibnamefont {Bromberger}}, \bibinfo
  {author} {\bibfnamefont {S.~B.}\ \bibnamefont {Wilkins}}, \bibinfo {author}
  {\bibfnamefont {V.}~\bibnamefont {Khanna}}, \bibinfo {author} {\bibfnamefont
  {A.~D.}\ \bibnamefont {Caviglia}}, \bibinfo {author} {\bibfnamefont {Y.-D.}\
  \bibnamefont {Chuang}}, \bibinfo {author} {\bibfnamefont {W.~S.}\
  \bibnamefont {Lee}}, \bibinfo {author} {\bibfnamefont {W.~F.}\ \bibnamefont
  {Schlotter}}, \bibinfo {author} {\bibfnamefont {J.~J.}\ \bibnamefont
  {Turner}}, \bibinfo {author} {\bibfnamefont {M.~P.}\ \bibnamefont {Minitti}},
  \bibinfo {author} {\bibfnamefont {O.}~\bibnamefont {Krupin}}, \bibinfo
  {author} {\bibfnamefont {Z.~J.}\ \bibnamefont {Xu}}, \bibinfo {author}
  {\bibfnamefont {J.~S.}\ \bibnamefont {Wen}}, \bibinfo {author} {\bibfnamefont
  {G.~D.}\ \bibnamefont {Gu}}, \bibinfo {author} {\bibfnamefont {S.~S.}\
  \bibnamefont {Dhesi}}, \bibinfo {author} {\bibfnamefont {A.}~\bibnamefont
  {Cavalleri}}, \ and\ \bibinfo {author} {\bibfnamefont {J.~P.}\ \bibnamefont
  {Hill}},\ }\href {\doibase 10.1103/PhysRevLett.112.157002} {\bibfield
  {journal} {\bibinfo  {journal} {Phys. Rev. Lett.}\ }\textbf {\bibinfo
  {volume} {112}},\ \bibinfo {pages} {157002} (\bibinfo {year}
  {2014})}\BibitemShut {NoStop}%
\bibitem [{\citenamefont {Khanna}\ \emph {et~al.}(2016)\citenamefont {Khanna},
  \citenamefont {Mankowsky}, \citenamefont {Petrich}, \citenamefont
  {Bromberger}, \citenamefont {Cavill}, \citenamefont {M\"ohr-Vorobeva},
  \citenamefont {Nicoletti}, \citenamefont {Laplace}, \citenamefont {Gu},
  \citenamefont {Hill}, \citenamefont {F{\"o}rst}, \citenamefont {Cavalleri},\
  and\ \citenamefont {Dhesi}}]{Khanna2016}%
  \BibitemOpen
  \bibfield  {author} {\bibinfo {author} {\bibfnamefont {V.}~\bibnamefont
  {Khanna}}, \bibinfo {author} {\bibfnamefont {R.}~\bibnamefont {Mankowsky}},
  \bibinfo {author} {\bibfnamefont {M.}~\bibnamefont {Petrich}}, \bibinfo
  {author} {\bibfnamefont {H.}~\bibnamefont {Bromberger}}, \bibinfo {author}
  {\bibfnamefont {S.~A.}\ \bibnamefont {Cavill}}, \bibinfo {author}
  {\bibfnamefont {E.}~\bibnamefont {M\"ohr-Vorobeva}}, \bibinfo {author}
  {\bibfnamefont {D.}~\bibnamefont {Nicoletti}}, \bibinfo {author}
  {\bibfnamefont {Y.}~\bibnamefont {Laplace}}, \bibinfo {author} {\bibfnamefont
  {G.~D.}\ \bibnamefont {Gu}}, \bibinfo {author} {\bibfnamefont {J.~P.}\
  \bibnamefont {Hill}}, \bibinfo {author} {\bibfnamefont {M.}~\bibnamefont
  {F{\"o}rst}}, \bibinfo {author} {\bibfnamefont {A.}~\bibnamefont
  {Cavalleri}}, \ and\ \bibinfo {author} {\bibfnamefont {S.~S.}\ \bibnamefont
  {Dhesi}},\ }\href {\doibase 10.1103/PhysRevB.93.224522} {\bibfield  {journal}
  {\bibinfo  {journal} {Phys. Rev. B}\ }\textbf {\bibinfo {volume} {93}},\
  \bibinfo {pages} {224522} (\bibinfo {year} {2016})}\BibitemShut {NoStop}%
\bibitem [{\citenamefont {Abbamonte}\ \emph {et~al.}(2005)\citenamefont
  {Abbamonte}, \citenamefont {Rusydi}, \citenamefont {Smadici}, \citenamefont
  {Gu}, \citenamefont {Sawatzky},\ and\ \citenamefont {Feng}}]{Abbamonte2005}%
  \BibitemOpen
  \bibfield  {author} {\bibinfo {author} {\bibfnamefont {P.}~\bibnamefont
  {Abbamonte}}, \bibinfo {author} {\bibfnamefont {A.}~\bibnamefont {Rusydi}},
  \bibinfo {author} {\bibfnamefont {S.}~\bibnamefont {Smadici}}, \bibinfo
  {author} {\bibfnamefont {G.~D.}\ \bibnamefont {Gu}}, \bibinfo {author}
  {\bibfnamefont {G.~A.}\ \bibnamefont {Sawatzky}}, \ and\ \bibinfo {author}
  {\bibfnamefont {D.~L.}\ \bibnamefont {Feng}},\ }\href {\doibase
  10.1038/nphys178} {\bibfield  {journal} {\bibinfo  {journal} {Nature
  Physics}\ }\textbf {\bibinfo {volume} {1}},\ \bibinfo {pages} {155} (\bibinfo
  {year} {2005})}\BibitemShut {NoStop}%
\bibitem [{\citenamefont {H\"ucker}\ \emph {et~al.}(2011)\citenamefont
  {H\"ucker}, \citenamefont {v.~Zimmermann}, \citenamefont {Gu}, \citenamefont
  {Xu}, \citenamefont {Wen}, \citenamefont {Xu}, \citenamefont {Kang},
  \citenamefont {Zheludev},\ and\ \citenamefont {Tranquada}}]{Hucker2011}%
  \BibitemOpen
  \bibfield  {author} {\bibinfo {author} {\bibfnamefont {M.}~\bibnamefont
  {H\"ucker}}, \bibinfo {author} {\bibfnamefont {M.}~\bibnamefont
  {v.~Zimmermann}}, \bibinfo {author} {\bibfnamefont {G.~D.}\ \bibnamefont
  {Gu}}, \bibinfo {author} {\bibfnamefont {Z.~J.}\ \bibnamefont {Xu}}, \bibinfo
  {author} {\bibfnamefont {J.~S.}\ \bibnamefont {Wen}}, \bibinfo {author}
  {\bibfnamefont {G.}~\bibnamefont {Xu}}, \bibinfo {author} {\bibfnamefont
  {H.~J.}\ \bibnamefont {Kang}}, \bibinfo {author} {\bibfnamefont
  {A.}~\bibnamefont {Zheludev}}, \ and\ \bibinfo {author} {\bibfnamefont
  {J.~M.}\ \bibnamefont {Tranquada}},\ }\href {\doibase
  10.1103/PhysRevB.83.104506} {\bibfield  {journal} {\bibinfo  {journal} {Phys.
  Rev. B}\ }\textbf {\bibinfo {volume} {83}},\ \bibinfo {pages} {104506}
  (\bibinfo {year} {2011})}\BibitemShut {NoStop}%
\bibitem [{\citenamefont {Tranquada}\ \emph {et~al.}(2004)\citenamefont
  {Tranquada}, \citenamefont {Woo}, \citenamefont {Perring}, \citenamefont
  {Goka}, \citenamefont {Gu}, \citenamefont {Xu}, \citenamefont {Fujita},\ and\
  \citenamefont {Yamada}}]{Tranquada2004}%
  \BibitemOpen
  \bibfield  {author} {\bibinfo {author} {\bibfnamefont {J.~M.}\ \bibnamefont
  {Tranquada}}, \bibinfo {author} {\bibfnamefont {H.}~\bibnamefont {Woo}},
  \bibinfo {author} {\bibfnamefont {T.~G.}\ \bibnamefont {Perring}}, \bibinfo
  {author} {\bibfnamefont {H.}~\bibnamefont {Goka}}, \bibinfo {author}
  {\bibfnamefont {G.~D.}\ \bibnamefont {Gu}}, \bibinfo {author} {\bibfnamefont
  {G.}~\bibnamefont {Xu}}, \bibinfo {author} {\bibfnamefont {M.}~\bibnamefont
  {Fujita}}, \ and\ \bibinfo {author} {\bibfnamefont {K.}~\bibnamefont
  {Yamada}},\ }\href {\doibase 10.1038/nature02574} {\bibfield  {journal}
  {\bibinfo  {journal} {Nature}\ }\textbf {\bibinfo {volume} {429}},\ \bibinfo
  {pages} {534} (\bibinfo {year} {2004})}\BibitemShut {NoStop}%
\bibitem [{\citenamefont {Mitrano}\ \emph {et~al.}(2019)\citenamefont
  {Mitrano}, \citenamefont {Lee}, \citenamefont {Husain}, \citenamefont
  {Delacretaz}, \citenamefont {Zhu}, \citenamefont {de~la Pe{\~n}a~Munoz},
  \citenamefont {Sun}, \citenamefont {Joe}, \citenamefont {Reid}, \citenamefont
  {Wandel}, \citenamefont {Coslovich}, \citenamefont {Schlotter}, \citenamefont
  {van Driel}, \citenamefont {Schneeloch}, \citenamefont {Gu}, \citenamefont
  {Hartnoll}, \citenamefont {Goldenfeld},\ and\ \citenamefont
  {Abbamonte}}]{Mitrano2018}%
  \BibitemOpen
  \bibfield  {author} {\bibinfo {author} {\bibfnamefont {M.}~\bibnamefont
  {Mitrano}}, \bibinfo {author} {\bibfnamefont {S.}~\bibnamefont {Lee}},
  \bibinfo {author} {\bibfnamefont {A.~A.}\ \bibnamefont {Husain}}, \bibinfo
  {author} {\bibfnamefont {L.}~\bibnamefont {Delacretaz}}, \bibinfo {author}
  {\bibfnamefont {M.}~\bibnamefont {Zhu}}, \bibinfo {author} {\bibfnamefont
  {G.}~\bibnamefont {de~la Pe{\~n}a~Munoz}}, \bibinfo {author} {\bibfnamefont
  {S.~X.-L.}\ \bibnamefont {Sun}}, \bibinfo {author} {\bibfnamefont {Y.~I.}\
  \bibnamefont {Joe}}, \bibinfo {author} {\bibfnamefont {A.~H.}\ \bibnamefont
  {Reid}}, \bibinfo {author} {\bibfnamefont {S.~F.}\ \bibnamefont {Wandel}},
  \bibinfo {author} {\bibfnamefont {G.}~\bibnamefont {Coslovich}}, \bibinfo
  {author} {\bibfnamefont {W.}~\bibnamefont {Schlotter}}, \bibinfo {author}
  {\bibfnamefont {T.}~\bibnamefont {van Driel}}, \bibinfo {author}
  {\bibfnamefont {J.}~\bibnamefont {Schneeloch}}, \bibinfo {author}
  {\bibfnamefont {G.~D.}\ \bibnamefont {Gu}}, \bibinfo {author} {\bibfnamefont
  {S.}~\bibnamefont {Hartnoll}}, \bibinfo {author} {\bibfnamefont
  {N.}~\bibnamefont {Goldenfeld}}, \ and\ \bibinfo {author} {\bibfnamefont
  {P.}~\bibnamefont {Abbamonte}},\ }\href {\doibase 10.1126/sciadv.aax3346}
  {\bibfield  {journal} {\bibinfo  {journal} {Science Advances}\ }\textbf
  {\bibinfo {volume} {5}},\ \bibinfo {pages} {eaax3346} (\bibinfo {year}
  {2019})}\BibitemShut {NoStop}%
\bibitem [{\citenamefont {Schlotter}\ \emph {et~al.}(2012)\citenamefont
  {Schlotter}, \citenamefont {Turner}, \citenamefont {Rowen}, \citenamefont
  {Heimann}, \citenamefont {Holmes}, \citenamefont {Krupin}, \citenamefont
  {Messerschmidt}, \citenamefont {Moeller}, \citenamefont {Krzywinski},
  \citenamefont {Soufli}, \citenamefont {Fern\'{a}ndez-Perea}, \citenamefont
  {Kelez}, \citenamefont {Lee}, \citenamefont {Coffee}, \citenamefont {Hays},
  \citenamefont {Beye}, \citenamefont {Gerken}, \citenamefont {Sorgenfrei},
  \citenamefont {Hau-Riege}, \citenamefont {Juha}, \citenamefont {Chalupsky},
  \citenamefont {Hajkova}, \citenamefont {Mancuso}, \citenamefont {Singer},
  \citenamefont {Yefanov}, \citenamefont {Vartanyants}, \citenamefont
  {Cadenazzi}, \citenamefont {Abbey}, \citenamefont {Nugent}, \citenamefont
  {Sinn}, \citenamefont {L{\"u}ning}, \citenamefont {Schaffert}, \citenamefont
  {Eisebitt}, \citenamefont {Lee}, \citenamefont {Scherz}, \citenamefont
  {Nilsson},\ and\ \citenamefont {Wurth}}]{Schlotter2012}%
  \BibitemOpen
  \bibfield  {author} {\bibinfo {author} {\bibfnamefont {W.~F.}\ \bibnamefont
  {Schlotter}}, \bibinfo {author} {\bibfnamefont {J.~J.}\ \bibnamefont
  {Turner}}, \bibinfo {author} {\bibfnamefont {M.}~\bibnamefont {Rowen}},
  \bibinfo {author} {\bibfnamefont {P.}~\bibnamefont {Heimann}}, \bibinfo
  {author} {\bibfnamefont {M.}~\bibnamefont {Holmes}}, \bibinfo {author}
  {\bibfnamefont {O.}~\bibnamefont {Krupin}}, \bibinfo {author} {\bibfnamefont
  {M.}~\bibnamefont {Messerschmidt}}, \bibinfo {author} {\bibfnamefont
  {S.}~\bibnamefont {Moeller}}, \bibinfo {author} {\bibfnamefont
  {J.}~\bibnamefont {Krzywinski}}, \bibinfo {author} {\bibfnamefont
  {R.}~\bibnamefont {Soufli}}, \bibinfo {author} {\bibfnamefont
  {M.}~\bibnamefont {Fern\'{a}ndez-Perea}}, \bibinfo {author} {\bibfnamefont
  {N.}~\bibnamefont {Kelez}}, \bibinfo {author} {\bibfnamefont
  {S.}~\bibnamefont {Lee}}, \bibinfo {author} {\bibfnamefont {R.}~\bibnamefont
  {Coffee}}, \bibinfo {author} {\bibfnamefont {G.}~\bibnamefont {Hays}},
  \bibinfo {author} {\bibfnamefont {M.}~\bibnamefont {Beye}}, \bibinfo {author}
  {\bibfnamefont {N.}~\bibnamefont {Gerken}}, \bibinfo {author} {\bibfnamefont
  {F.}~\bibnamefont {Sorgenfrei}}, \bibinfo {author} {\bibfnamefont
  {S.}~\bibnamefont {Hau-Riege}}, \bibinfo {author} {\bibfnamefont
  {L.}~\bibnamefont {Juha}}, \bibinfo {author} {\bibfnamefont {J.}~\bibnamefont
  {Chalupsky}}, \bibinfo {author} {\bibfnamefont {V.}~\bibnamefont {Hajkova}},
  \bibinfo {author} {\bibfnamefont {A.~P.}\ \bibnamefont {Mancuso}}, \bibinfo
  {author} {\bibfnamefont {A.}~\bibnamefont {Singer}}, \bibinfo {author}
  {\bibfnamefont {O.}~\bibnamefont {Yefanov}}, \bibinfo {author} {\bibfnamefont
  {I.~A.}\ \bibnamefont {Vartanyants}}, \bibinfo {author} {\bibfnamefont
  {G.}~\bibnamefont {Cadenazzi}}, \bibinfo {author} {\bibfnamefont
  {B.}~\bibnamefont {Abbey}}, \bibinfo {author} {\bibfnamefont {K.~A.}\
  \bibnamefont {Nugent}}, \bibinfo {author} {\bibfnamefont {H.}~\bibnamefont
  {Sinn}}, \bibinfo {author} {\bibfnamefont {J.}~\bibnamefont {L{\"u}ning}},
  \bibinfo {author} {\bibfnamefont {S.}~\bibnamefont {Schaffert}}, \bibinfo
  {author} {\bibfnamefont {S.}~\bibnamefont {Eisebitt}}, \bibinfo {author}
  {\bibfnamefont {W.~S.}\ \bibnamefont {Lee}}, \bibinfo {author} {\bibfnamefont
  {A.}~\bibnamefont {Scherz}}, \bibinfo {author} {\bibfnamefont {A.~R.}\
  \bibnamefont {Nilsson}}, \ and\ \bibinfo {author} {\bibfnamefont
  {W.}~\bibnamefont {Wurth}},\ }\href {\doibase 10.1063/1.3698294} {\bibfield
  {journal} {\bibinfo  {journal} {Review of Scientific Instruments}\ }\textbf
  {\bibinfo {volume} {83}},\ \bibinfo {pages} {043107} (\bibinfo {year}
  {2012})}\BibitemShut {NoStop}%
\bibitem [{\citenamefont {Doering}\ \emph {et~al.}(2011)\citenamefont
  {Doering}, \citenamefont {Chuang}, \citenamefont {Andresen}, \citenamefont
  {Chow}, \citenamefont {Contarato}, \citenamefont {Cummings}, \citenamefont
  {Domning}, \citenamefont {Joseph}, \citenamefont {Pepper}, \citenamefont
  {Smith}, \citenamefont {Zizka}, \citenamefont {Ford}, \citenamefont {Lee},
  \citenamefont {Weaver}, \citenamefont {Patthey}, \citenamefont {Weizeorick},
  \citenamefont {Hussain},\ and\ \citenamefont {Denes}}]{Doering2011}%
  \BibitemOpen
  \bibfield  {author} {\bibinfo {author} {\bibfnamefont {D.}~\bibnamefont
  {Doering}}, \bibinfo {author} {\bibfnamefont {Y.~D.}\ \bibnamefont {Chuang}},
  \bibinfo {author} {\bibfnamefont {N.}~\bibnamefont {Andresen}}, \bibinfo
  {author} {\bibfnamefont {K.}~\bibnamefont {Chow}}, \bibinfo {author}
  {\bibfnamefont {D.}~\bibnamefont {Contarato}}, \bibinfo {author}
  {\bibfnamefont {C.}~\bibnamefont {Cummings}}, \bibinfo {author}
  {\bibfnamefont {E.}~\bibnamefont {Domning}}, \bibinfo {author} {\bibfnamefont
  {J.}~\bibnamefont {Joseph}}, \bibinfo {author} {\bibfnamefont {J.~S.}\
  \bibnamefont {Pepper}}, \bibinfo {author} {\bibfnamefont {B.}~\bibnamefont
  {Smith}}, \bibinfo {author} {\bibfnamefont {G.}~\bibnamefont {Zizka}},
  \bibinfo {author} {\bibfnamefont {C.}~\bibnamefont {Ford}}, \bibinfo {author}
  {\bibfnamefont {W.~S.}\ \bibnamefont {Lee}}, \bibinfo {author} {\bibfnamefont
  {M.}~\bibnamefont {Weaver}}, \bibinfo {author} {\bibfnamefont
  {L.}~\bibnamefont {Patthey}}, \bibinfo {author} {\bibfnamefont
  {J.}~\bibnamefont {Weizeorick}}, \bibinfo {author} {\bibfnamefont
  {Z.}~\bibnamefont {Hussain}}, \ and\ \bibinfo {author} {\bibfnamefont
  {P.}~\bibnamefont {Denes}},\ }\href {\doibase 10.1063/1.3609862} {\bibfield
  {journal} {\bibinfo  {journal} {Review of Scientific Instruments}\ }\textbf
  {\bibinfo {volume} {82}},\ \bibinfo {pages} {073303} (\bibinfo {year}
  {2011})}\BibitemShut {NoStop}%
\bibitem [{\citenamefont {Lemke}\ \emph {et~al.}(2013)\citenamefont {Lemke},
  \citenamefont {Weaver}, \citenamefont {Chollet}, \citenamefont {Robinson},
  \citenamefont {Glownia}, \citenamefont {Zhu}, \citenamefont {Bionta},
  \citenamefont {Cammarata}, \citenamefont {Harmand}, \citenamefont {Coffee},\
  and\ \citenamefont {Fritz}}]{Lemke2013}%
  \BibitemOpen
  \bibfield  {author} {\bibinfo {author} {\bibfnamefont {H.~T.}\ \bibnamefont
  {Lemke}}, \bibinfo {author} {\bibfnamefont {M.}~\bibnamefont {Weaver}},
  \bibinfo {author} {\bibfnamefont {M.}~\bibnamefont {Chollet}}, \bibinfo
  {author} {\bibfnamefont {J.}~\bibnamefont {Robinson}}, \bibinfo {author}
  {\bibfnamefont {J.~M.}\ \bibnamefont {Glownia}}, \bibinfo {author}
  {\bibfnamefont {D.}~\bibnamefont {Zhu}}, \bibinfo {author} {\bibfnamefont
  {M.~R.}\ \bibnamefont {Bionta}}, \bibinfo {author} {\bibfnamefont
  {M.}~\bibnamefont {Cammarata}}, \bibinfo {author} {\bibfnamefont
  {M.}~\bibnamefont {Harmand}}, \bibinfo {author} {\bibfnamefont {R.~N.}\
  \bibnamefont {Coffee}}, \ and\ \bibinfo {author} {\bibfnamefont {D.~M.}\
  \bibnamefont {Fritz}},\ }in\ \href {\doibase 10.1117/12.2017603} {\emph
  {\bibinfo {booktitle} {SPIE Optics + Optoelectronics}}},\ \bibinfo {editor}
  {edited by\ \bibinfo {editor} {\bibfnamefont {T.}~\bibnamefont
  {Tschentscher}}\ and\ \bibinfo {editor} {\bibfnamefont {K.}~\bibnamefont
  {Tiedtke}}}\ (\bibinfo  {publisher} {SPIE},\ \bibinfo {year} {2013})\ pp.\
  \bibinfo {pages} {87780S--5}\BibitemShut {NoStop}%
\bibitem [{\citenamefont {Harmand}\ \emph {et~al.}(2013)\citenamefont
  {Harmand}, \citenamefont {Coffee}, \citenamefont {Bionta}, \citenamefont
  {Chollet}, \citenamefont {French}, \citenamefont {Zhu}, \citenamefont
  {Fritz}, \citenamefont {Lemke}, \citenamefont {Medvedev}, \citenamefont
  {Ziaja}, \citenamefont {Toleikis},\ and\ \citenamefont
  {Cammarata}}]{Harmand2013}%
  \BibitemOpen
  \bibfield  {author} {\bibinfo {author} {\bibfnamefont {M.}~\bibnamefont
  {Harmand}}, \bibinfo {author} {\bibfnamefont {R.}~\bibnamefont {Coffee}},
  \bibinfo {author} {\bibfnamefont {M.~R.}\ \bibnamefont {Bionta}}, \bibinfo
  {author} {\bibfnamefont {M.}~\bibnamefont {Chollet}}, \bibinfo {author}
  {\bibfnamefont {D.}~\bibnamefont {French}}, \bibinfo {author} {\bibfnamefont
  {D.}~\bibnamefont {Zhu}}, \bibinfo {author} {\bibfnamefont {D.~M.}\
  \bibnamefont {Fritz}}, \bibinfo {author} {\bibfnamefont {H.~T.}\ \bibnamefont
  {Lemke}}, \bibinfo {author} {\bibfnamefont {N.}~\bibnamefont {Medvedev}},
  \bibinfo {author} {\bibfnamefont {B.}~\bibnamefont {Ziaja}}, \bibinfo
  {author} {\bibfnamefont {S.}~\bibnamefont {Toleikis}}, \ and\ \bibinfo
  {author} {\bibfnamefont {M.}~\bibnamefont {Cammarata}},\ }\href {\doibase
  10.1038/nphoton.2013.11} {\bibfield  {journal} {\bibinfo  {journal} {Nature
  Photonics}\ }\textbf {\bibinfo {volume} {7}},\ \bibinfo {pages} {215}
  (\bibinfo {year} {2013})}\BibitemShut {NoStop}%
\bibitem [{\citenamefont {Chuang}\ \emph {et~al.}(2017)\citenamefont {Chuang},
  \citenamefont {Shao}, \citenamefont {Cruz}, \citenamefont {Hanzel},
  \citenamefont {Brown}, \citenamefont {Frano}, \citenamefont {Qiao},
  \citenamefont {Smith}, \citenamefont {Domning}, \citenamefont {Huang},
  \citenamefont {Wray}, \citenamefont {Lee}, \citenamefont {Shen},
  \citenamefont {Devereaux}, \citenamefont {Chiou}, \citenamefont {Pong},
  \citenamefont {Yashchuk}, \citenamefont {Gullikson}, \citenamefont
  {Reininger}, \citenamefont {Yang}, \citenamefont {Guo}, \citenamefont
  {Duarte},\ and\ \citenamefont {Hussain}}]{Chuang2017}%
  \BibitemOpen
  \bibfield  {author} {\bibinfo {author} {\bibfnamefont {Y.~D.}\ \bibnamefont
  {Chuang}}, \bibinfo {author} {\bibfnamefont {Y.~C.}\ \bibnamefont {Shao}},
  \bibinfo {author} {\bibfnamefont {A.}~\bibnamefont {Cruz}}, \bibinfo {author}
  {\bibfnamefont {K.}~\bibnamefont {Hanzel}}, \bibinfo {author} {\bibfnamefont
  {A.}~\bibnamefont {Brown}}, \bibinfo {author} {\bibfnamefont
  {A.}~\bibnamefont {Frano}}, \bibinfo {author} {\bibfnamefont
  {R.}~\bibnamefont {Qiao}}, \bibinfo {author} {\bibfnamefont {B.}~\bibnamefont
  {Smith}}, \bibinfo {author} {\bibfnamefont {E.}~\bibnamefont {Domning}},
  \bibinfo {author} {\bibfnamefont {S.~W.}\ \bibnamefont {Huang}}, \bibinfo
  {author} {\bibfnamefont {L.~A.}\ \bibnamefont {Wray}}, \bibinfo {author}
  {\bibfnamefont {W.~S.}\ \bibnamefont {Lee}}, \bibinfo {author} {\bibfnamefont
  {Z.-X.}\ \bibnamefont {Shen}}, \bibinfo {author} {\bibfnamefont {T.~P.}\
  \bibnamefont {Devereaux}}, \bibinfo {author} {\bibfnamefont {J.-W.}\
  \bibnamefont {Chiou}}, \bibinfo {author} {\bibfnamefont {W.-F.}\ \bibnamefont
  {Pong}}, \bibinfo {author} {\bibfnamefont {V.~V.}\ \bibnamefont {Yashchuk}},
  \bibinfo {author} {\bibfnamefont {E.}~\bibnamefont {Gullikson}}, \bibinfo
  {author} {\bibfnamefont {R.}~\bibnamefont {Reininger}}, \bibinfo {author}
  {\bibfnamefont {W.}~\bibnamefont {Yang}}, \bibinfo {author} {\bibfnamefont
  {J.}~\bibnamefont {Guo}}, \bibinfo {author} {\bibfnamefont {R.}~\bibnamefont
  {Duarte}}, \ and\ \bibinfo {author} {\bibfnamefont {Z.}~\bibnamefont
  {Hussain}},\ }\href {\doibase 10.1063/1.4974356} {\bibfield  {journal}
  {\bibinfo  {journal} {Review of Scientific Instruments}\ }\textbf {\bibinfo
  {volume} {88}},\ \bibinfo {pages} {013110} (\bibinfo {year}
  {2017})}\BibitemShut {NoStop}%
\bibitem [{\citenamefont {Ghiringhelli}\ \emph {et~al.}(2004)\citenamefont
  {Ghiringhelli}, \citenamefont {Brookes}, \citenamefont {Annese},
  \citenamefont {Berger}, \citenamefont {Dallera}, \citenamefont {Grioni},
  \citenamefont {Perfetti}, \citenamefont {Tagliaferri},\ and\ \citenamefont
  {Braicovich}}]{Ghiringhelli2004}%
  \BibitemOpen
  \bibfield  {author} {\bibinfo {author} {\bibfnamefont {G.}~\bibnamefont
  {Ghiringhelli}}, \bibinfo {author} {\bibfnamefont {N.~B.}\ \bibnamefont
  {Brookes}}, \bibinfo {author} {\bibfnamefont {E.}~\bibnamefont {Annese}},
  \bibinfo {author} {\bibfnamefont {H.}~\bibnamefont {Berger}}, \bibinfo
  {author} {\bibfnamefont {C.}~\bibnamefont {Dallera}}, \bibinfo {author}
  {\bibfnamefont {M.}~\bibnamefont {Grioni}}, \bibinfo {author} {\bibfnamefont
  {L.}~\bibnamefont {Perfetti}}, \bibinfo {author} {\bibfnamefont
  {A.}~\bibnamefont {Tagliaferri}}, \ and\ \bibinfo {author} {\bibfnamefont
  {L.}~\bibnamefont {Braicovich}},\ }\href {\doibase
  10.1103/PhysRevLett.92.117406} {\bibfield  {journal} {\bibinfo  {journal}
  {Phys. Rev. Lett.}\ }\textbf {\bibinfo {volume} {92}},\ \bibinfo {pages}
  {117406} (\bibinfo {year} {2004})}\BibitemShut {NoStop}%
\bibitem [{\citenamefont {Smadici}\ \emph {et~al.}(2013)\citenamefont
  {Smadici}, \citenamefont {Lee}, \citenamefont {Logvenov}, \citenamefont
  {Bozovic},\ and\ \citenamefont {Abbamonte}}]{Smadici2013}%
  \BibitemOpen
  \bibfield  {author} {\bibinfo {author} {\bibfnamefont {S.}~\bibnamefont
  {Smadici}}, \bibinfo {author} {\bibfnamefont {J.~C.~T.}\ \bibnamefont {Lee}},
  \bibinfo {author} {\bibfnamefont {G.}~\bibnamefont {Logvenov}}, \bibinfo
  {author} {\bibfnamefont {I.}~\bibnamefont {Bozovic}}, \ and\ \bibinfo
  {author} {\bibfnamefont {P.}~\bibnamefont {Abbamonte}},\ }\href {\doibase
  10.1088/0953-8984/26/2/025303} {\bibfield  {journal} {\bibinfo  {journal} {J.
  Phys.: Condens. Matter}\ }\textbf {\bibinfo {volume} {26}},\ \bibinfo {pages}
  {025303} (\bibinfo {year} {2013})}\BibitemShut {NoStop}%
\bibitem [{\citenamefont {Achkar}\ \emph {et~al.}(2016)\citenamefont {Achkar},
  \citenamefont {Zwiebler}, \citenamefont {McMahon}, \citenamefont {He},
  \citenamefont {Sutarto}, \citenamefont {Djianto}, \citenamefont {Hao},
  \citenamefont {Gingras}, \citenamefont {H{\"u}cker}, \citenamefont {Gu},
  \citenamefont {Revcolevschi}, \citenamefont {Zhang}, \citenamefont {Kim},
  \citenamefont {Geck},\ and\ \citenamefont {Hawthorn}}]{Achkar2016}%
  \BibitemOpen
  \bibfield  {author} {\bibinfo {author} {\bibfnamefont {A.~J.}\ \bibnamefont
  {Achkar}}, \bibinfo {author} {\bibfnamefont {M.}~\bibnamefont {Zwiebler}},
  \bibinfo {author} {\bibfnamefont {C.}~\bibnamefont {McMahon}}, \bibinfo
  {author} {\bibfnamefont {F.}~\bibnamefont {He}}, \bibinfo {author}
  {\bibfnamefont {R.}~\bibnamefont {Sutarto}}, \bibinfo {author} {\bibfnamefont
  {I.}~\bibnamefont {Djianto}}, \bibinfo {author} {\bibfnamefont
  {Z.}~\bibnamefont {Hao}}, \bibinfo {author} {\bibfnamefont {M.~J.~P.}\
  \bibnamefont {Gingras}}, \bibinfo {author} {\bibfnamefont {M.}~\bibnamefont
  {H{\"u}cker}}, \bibinfo {author} {\bibfnamefont {G.~D.}\ \bibnamefont {Gu}},
  \bibinfo {author} {\bibfnamefont {A.}~\bibnamefont {Revcolevschi}}, \bibinfo
  {author} {\bibfnamefont {H.}~\bibnamefont {Zhang}}, \bibinfo {author}
  {\bibfnamefont {Y.~J.}\ \bibnamefont {Kim}}, \bibinfo {author} {\bibfnamefont
  {J.}~\bibnamefont {Geck}}, \ and\ \bibinfo {author} {\bibfnamefont {D.~G.}\
  \bibnamefont {Hawthorn}},\ }\href@noop {} {\bibfield  {journal} {\bibinfo
  {journal} {Science}\ }\textbf {\bibinfo {volume} {351}},\ \bibinfo {pages}
  {576} (\bibinfo {year} {2016})}\BibitemShut {NoStop}%
\bibitem [{\citenamefont {Coppersmith}\ and\ \citenamefont
  {Varma}(1984)}]{Coppersmith1984}%
  \BibitemOpen
  \bibfield  {author} {\bibinfo {author} {\bibfnamefont {S.~N.}\ \bibnamefont
  {Coppersmith}}\ and\ \bibinfo {author} {\bibfnamefont {C.~M.}\ \bibnamefont
  {Varma}},\ }\href {\doibase 10.1103/PhysRevB.30.3566} {\bibfield  {journal}
  {\bibinfo  {journal} {Phys. Rev. B}\ }\textbf {\bibinfo {volume} {30}},\
  \bibinfo {pages} {3566} (\bibinfo {year} {1984})}\BibitemShut {NoStop}%
\bibitem [{\citenamefont {Valla}\ \emph {et~al.}(2006)\citenamefont {Valla},
  \citenamefont {Fedorov}, \citenamefont {Lee}, \citenamefont {Davis},\ and\
  \citenamefont {Gu}}]{Valla2006}%
  \BibitemOpen
  \bibfield  {author} {\bibinfo {author} {\bibfnamefont {T.}~\bibnamefont
  {Valla}}, \bibinfo {author} {\bibfnamefont {A.~V.}\ \bibnamefont {Fedorov}},
  \bibinfo {author} {\bibfnamefont {J.}~\bibnamefont {Lee}}, \bibinfo {author}
  {\bibfnamefont {J.~C.}\ \bibnamefont {Davis}}, \ and\ \bibinfo {author}
  {\bibfnamefont {G.~D.}\ \bibnamefont {Gu}},\ }\href {\doibase
  10.1126/science.1134742} {\bibfield  {journal} {\bibinfo  {journal}
  {Science}\ }\textbf {\bibinfo {volume} {314}},\ \bibinfo {pages} {1914}
  (\bibinfo {year} {2006})}\BibitemShut {NoStop}%
\end{thebibliography}%

\end{document}